\documentclass[10pt,journal,a4paper,draftcls,oneside,onecolumn]{IEEEtran}
\usepackage{times} 
\usepackage{comment}
\usepackage{csquotes}
\usepackage{graphicx}
\usepackage{subfigure}
\usepackage{cite}
\usepackage{citesort}
\usepackage{color}
\usepackage{psfrag}
\usepackage{amssymb}
\usepackage{epsfig}
\usepackage{pifont}
\usepackage{amsmath}
\usepackage{bm}
\usepackage{mathtools}
\usepackage{array}
\usepackage{multicol}
\usepackage[T1]{fontenc}
\usepackage{makecell}
\usepackage{algorithm}
\usepackage{algorithmic}
\usepackage{xcolor}

\newcommand{\fref}[1]{Fig.~\ref{#1}}
\newcommand{\tref}[1]{Table~\ref{#1}} 
\newcommand{\alref}[1]{Algorithm~\ref{#1}}
\newcommand{\apref}[1]{Appendix~\ref{#1}}
\makeatother 

\begin{document}
\title{Constant-Weight Sparse Vector Coded  Augmented-AFDM}

\author{\IEEEauthorblockN{Zhou Lu, Mohammed El-Hajjar, \IEEEmembership{Senior Member,~IEEE,} and Lie-liang Yang, \IEEEmembership{Fellow,~IEEE, }}}
\maketitle

\begin{abstract}
Sparse vector coding (SVC) and its variants convey information through sparse vectors (SVs), or sparse support patterns and detecting information by solving sparse recovery problems, making them attractive for short-packet ultra-reliable low-latency communications (URLLC). However, as the required information rate increases, conventional SVC schemes or the enhanced SVC (ESVC) schemes require a rapidly growing number of SVs, resulting in significant increase of complexity and receiver processing latency. On the other hand, when communicating in doubly selective channels, orthogonal frequency division multiplexing (OFDM) suffers from the Doppler spread induced inter-carrier interference, while affine frequency division multiplexing (AFDM) experiences the same peak-to-average power ratio (PAPR) problem as OFDM, although it allows diversity gain from channel's double selectivity. To address these issues, this paper proposes a constant-weight sparse vector coded augmented AFDM (CW-SVC/A$^2$FDM ) scheme, along with a corresponding joint support and symbol detection algorithm. In this scheme, SVs are represented by constant-weight codewords that feature a fixed Hamming weight and a large Hamming distance, resulting in relatively small SV sets. Consequently, highly reliable sparse recovery is achievable. Meanwhile, because the primary information carrier shifts from SV candidates to channel-coded quadrature amplitude modulation (QAM) symbols, the proposed detection algorithm delivers both highly reliable information detection and low-latency receiver processing. Simulation results compare various SVC/ESVC schemes and diverse coding schemes employed by the CW-SVC/A$^2$FDM, including Hamming, BCH, RS, and polar codes, for QAM symbol protection. The results demonstrate that the proposed scheme provides a distinct reliability advantage over conventional SVC/ESVC schemes, achieving a gain of up to 14 dB. In the CW-SVC/A$^2$FDM system, errors are dominated by QAM demodulation, with polar codes offering the strongest protection among the evaluated options. Furthermore, the study indicates that the CWC's weight and minimum Hamming distance must be carefully selected to balance detection reliability, QAM modulation order, and sparse recovery feasibility, thereby maximizing the overall system's performance.
\end{abstract}


\section{Introduction}\label{section-ESVC-Introduction}

Ultra-reliable low-latency communications (URLLC) and massive machine-type communications (mMTC) represent two distinct service categories in modern mobile communication networks, each characterized by different communication requirements \cite{3gpp38.913}. URLLC is designed to support the applications that are highly sensitive to reliability and latency \cite{siddiqui2023urllc}, while mMTC targets mostly the large scale machine-to-machine communications \cite{goshi2024modeling}. Moreover, these two scenarios often share some overlapping application domains, including vehicle-to-vehicle (V2V) communications \cite{yang2020ultra}, autonomous driving \cite{osorio2022towards}, internet of things (IoT) \cite{nguyen20216g}, and remote sensing and control \cite{liu2022integrated}. For example, in autonomous driving, ensuring passenger safety and maintaining proper vehicle operation require highly reliable and low-latency communications both between vehicles and between vehicles and base stations.  These requirements are closely aligned with both mMTC and URLLC. Such applications typically require the rapid exchange of real-time data, including sensing information (such as temperature, humidity, traffic signals, and road conditions) and control commands (such as start/stop, acceleration/deceleration, and power switching).

Sparse vector coding (SVC) has emerged as a potential solution in URLLC by transmitting data through mapping information bits into sparse support patterns \cite{ji2018sparse}. Accordingly, its decoding process can be formulated as a sparse recovery problem under the compressed sensing (CS) framework \cite{zhang2022integrated}. Besides, CS also plays an important role in enabling low-overhead access and joint detection in mMTC \cite{alam2021survey}, thereby rendering the SVC scheme applicable in both URLLC and mMTC applications. To achieve better performance, in \cite{kim2020enhanced}, the authors proposed an enhanced SVC (ESVC) scheme, where information bits are jointly mapped to both the support patterns of the sparse vector and their conveyed QAM symbols. This design significantly reduces the number of the sparse support patterns required to represent the same amount of information in the SVC schemes, thereby enhancing both transmission efficiency and overall system performance. Building upon this, the authors of \cite{zhang2021sparse} introduced another novel short-packet transmission technique known as sparse superposition coding (SSC), which extends the ESVC framework by transmitting QAM symbols after pseudo-random spreading in the time-frequency domain. At the receiver, the decoding problem is reformulated as the detection of non-zero indices, which can be efficiently solved using the low-complexity energy detection algorithms. Many studies have demonstrated that SVC and its variants exhibit strong relevance in various application scenarios \cite{zhang2021sparse,yang2024multiple,zhang2023sparse,yang2023improved}. However, as the majority of information bits in the SVC and its variant schemes are still conveyed through the combinations of support patterns, a slight increase in the number of bits per symbol results in an exponential growth in the number of support patterns. This dramatically increases the computational complexity of the sparse recovery process at the receiver, leading to a significant rise in decoding latency, thereby posing challenges to meet the strict latency requirement of URLLC.

As a class of nonlinear codes, the codewords of a constant-weight code
(CWC) have the same Hamming
weight~\cite{ostergard2010classification}. Note that for binary vectors of an equal weight, the pairwise Hamming distance
  is always even, and increasing the weight of codewords does not by
  itself guarantee a larger minimum Hamming distance for an arbitrary
  CWC. For the constant-weight codebooks constructed in this paper,
  under the constraint that the codeword weight does not exceed half
  of the codeword length, the minimum Hamming distance is deliberately
  designed to increase with the codeword weight, thereby facilitating
  more reliable decoding. In contrast to activating random sparse
supports for conveying information in the conventional SVC scheme, the
CWC-derived sparse vector coding (CW-SVC) scheme yields less number of
sparse vector (SV) candidates but a relatively higher weight, which
results in a bigger number of supports per candidate SV. Consequently,
the primary information carrier shifts from SVs to QAM
symbols. Additionally, the use of higher code weights in CWC reduces
the total number of SV candidates, which decreases the computational
complexity required for the sparse recovery algorithms at the
receiver.

Moreover, high-reliability communications in high-mobility environments have been becoming essential in many applications, such as in V2V communications, autonomous driving and unmanned aerial vehicles (UAVs) \cite{el2020vehicle}. Classical multicarrier systems such as orthogonal frequency division multiplexing (OFDM) suffer from the inter-carrier interference (ICI) due to Doppler shifts. In contrast, affine frequency division multiplexing (AFDM), with its unique subcarrier waveform design, processes signals in the Doppler-delay domain and effectively mitigates Doppler-induced interference in addition to the delay-caused inter-symbol interference (ISI), thereby ensuring a robust performance in the doubly-selective fading scenarios \cite{bemani2023affine}. However, similar to OFDM, AFDM suffers from high peak-to-average power ratio (PAPR) problem, which poses challenges for implementing the size- and power-limited devices, such as that commonly found in IoT or autonomous driving systems \cite{kamal2025novel}. For these kinds of applications, a communication scheme with low PAPR while of high efficiency in the doubly-selective fading environments is critical.

To address these challenges, this paper proposes a
  novel CWC-aided ESVC, referred to as the CW-SVC scheme, for
  implementation with an augmented AFDM (A$^2$FDM) system
  \cite{11509294}, referred to as the CW-SVC/A$^2$FDM system.  The
  main novelties of the proposed scheme include the following. First,
  in the proposed CW-SVC/A$^2$FDM system, the support patterns of
  sparse vectors are formed by constant-weight codewords with large
  minimum Hamming distances, which shifts the primary information
  carrier of the system from the support patterns to the channel-coded
  QAM symbols. Second, with the FDTW method, the resulting
  constant-weight codebook turns support detection into a fixed-size
  candidate detection problem, which allows the candidate patterns to
  be detected in parallel and is, hence, suitable for low-latency
  applications. Third, the proposed scheme offers advantages in the
  transmission and detection of critical information, such as active
  users' identities in mMTC scenarios. In addition, the low PAPR of
  A$^2$FDM and its robustness in doubly selective fading environments
  can effectively enhance the practical utility of the CW-SVC/A$^2$FDM
  scheme, including drone communications in IoT networks and
  autonomous driving.

\begin{table*}
    \caption{Novelty comparison between our paper and the related references
      \cite{yang2024multiple,zhang2023sparse,ostergard2010classification,bemani2023affine,kamal2025novel,11509294,choi2026learning,sinha2024generalized}. }
    \label{tab:novelty}
    \centering
    \begin{tabular}{cccccccccc}\hline
         &  our paper&\cite{yang2024multiple}&  \cite{zhang2023sparse}  &\cite{ostergard2010classification}&  \cite{bemani2023affine}& \cite{kamal2025novel}&\cite{11509294}&\cite{choi2026learning}&\cite{sinha2024generalized}\\\hline
         A$^2$FDM & $\checkmark$ &&&&&&\checkmark&& \\ \hline
         Sparse vector coding&  $\checkmark$&$\checkmark$&  $\checkmark$  &&  &  &&&\\\hline
         Doubly-selective fading environment&  $\checkmark$&&    &&  $\checkmark$& &$\checkmark$&&\\\hline
         Constant weight coding&  $\checkmark$&&    &$\checkmark$  &  &  &&&\\ \hline
         Low PAPR& $\checkmark$&&   && &$\checkmark$ &$\checkmark$&& \\ \hline
        Linear coding for QAM& $\checkmark$&&   && & &&$\checkmark$ & \\ \hline
         Parallel detection& $\checkmark$&&   && & && &$\checkmark$ \\ \hline
    \end{tabular}
\end{table*}

The main contributions of the paper can be summarized as follow:

\begin{itemize}
  
\item We propose the CW-SVC scheme, where the support patterns are the codewords of a constant-weight code designed using the ``from dimension to weight (FDTW)'' construction~\cite{6979861}. The resulting large Hamming distance among the support patterns enables high-reliability discrimination during the sparse recovery process.
  
\item We propose amalgamating CW-SVC with A$^2$FDM to form the CW-SVC/A$^2$FDM system, where the QAM symbols conveyed on the supports are protected by conventional channel coding, yielding an enhanced system performance.
  
    \item We propose a practically effective joint support and symbol detection scheme, where the detection of different candidate patterns can be carried out in parallel to reduce the receiver's processing latency.
      
    \item For protection of the information conveyed by QAM symbols, various linear channel coding schemes are studied and compared, including Hamming, Reed–Solomon (RS), Bose–Chaudhuri–Hocquenghem (BCH), and polar codes. Simulation results demonstrate that while all coding schemes allow system performance enhancement, polar code exhibits the best performance among the coding schemes and parameter configurations evaluated in our simulations, enabling an up to 6.5~dB gain when compared to the case without using coding for protection of QAM symbols.
      
    \item Simulation results demonstrate that the proposed CW-SVC/A$^2$FDM system is capable of attaining a substantial improvement in terms of the overall performance. Compared with the conventional ESVC schemes, the proposed system exhibits a significantly enhanced reliability in SV detection, leading to the overall bit error rate (BER) primarily dominated by the errors arising from QAM demodulations, which can be protected by employing an appropriate channel coding scheme.
\end{itemize}

In Table \ref{tab:novelty}, we explicitly contrast our contributions with the state-of-the-art schemes in the literature.

The remainder of this paper is organized as follows: Section~\ref{section-ESVC-system model} describes the system model, including the CW-SVC scheme and the principles of A$^2$FDM. Section \ref{section-ESVC-CWC design} introduces the properties of CWC and describes the FDTW construction of CWC. Section \ref{section-ESVC-support detection} illustrates the proposed joint support and symbol detection, and compares the detection complexity and latency of the existing ESVC and proposed CW-SVC schemes. Then, in Section \ref{section-ESVC-performance and analysis}, performance results are presented to compare the performance of various schemes. Finally, in Section \ref{section-ESVC-Conclusion}, we conclude the research in this paper. An example of the CWC codewords used to obtain the results in this paper is presented in Appendix \ref{Appendix-CWC}.

\section{System Model}\label{section-ESVC-system model}

\begin{figure*}
    \centering
    \includegraphics[width=0.8\linewidth]{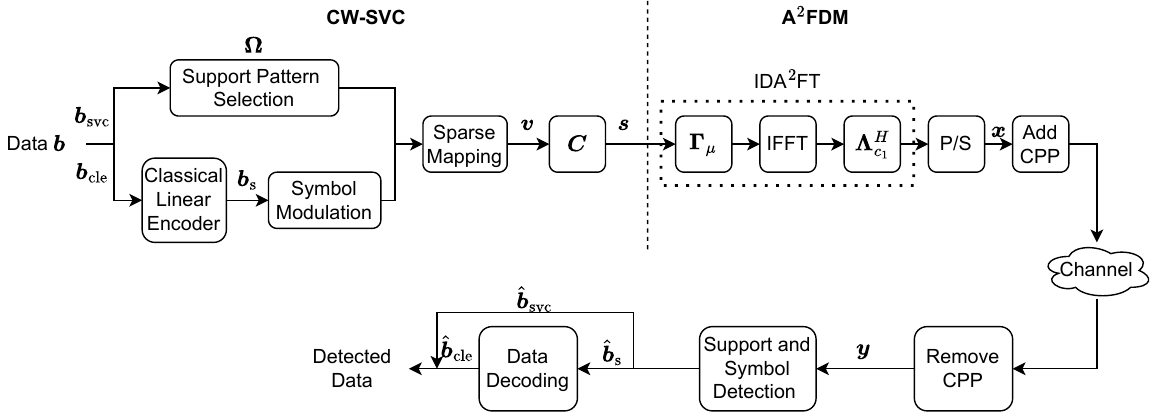}
    \caption{Transceiver diagram of the proposed single user system employing CW-SVC/A$^2$FDM.}
    \label{fig:system model}
\end{figure*}

The block diagram of the proposed CW-SVC/A$^2$FDM system is shown in \fref{fig:system model}, where the data stream $\bm{s}$ is constructed using the principle of the CW-SVC and transmitted via the A$^2$FDM framework \cite{11509294}. In this design, the information payload $\bm b$ is divided into two parts, $\bm b_\mathrm{svc}$ and $\bm
b_\mathrm{cle}$. $\bm b_\mathrm{svc}$ is used to select a support pattern from the codeword set formed by a CWC, while $\bm b_\mathrm{cle}$ is first processed by a classical linear encoder (CLE), which is then modulated into QAM symbols, followed by their mapping onto the supports of the selected pattern. The resulting sparse vector $\bm v$ is then compressed using matrix $\bm C$ to form the data stream $\bm
s$, which is matched to the number of subcarriers employed by the A$^2$FDM for further processing. As shown in \fref{fig:system model}, following the A$^2$FDM related processing, the signal is transmitted over a doubly selective fading channel. At the receiver side, support and symbol detection are performed by the proposed detection schemes designed based on the properties of CS and CWC.

It is worth noting that the proposed CW-SVC scheme is
  not limited to the A$^{2}$FDM waveform. The CW-SVC operation maps
  information bits to the constant-weight support patterns and to the
  QAM symbols before the waveform modulation. Hence, it can in
  principle be combined with various other linear modulation
  waveforms, including AFDM and OFDM. In this paper, A$^{2}$FDM is
  adopted as the transmission framework because of its robustness in
  doubly selective fading channels and its low PAPR. Therefore, the
  detection algorithms are presented in this context.

In further detail, the data stream $\bm b$ block with $B$ bits is separated into two sub-blocks of $\bm b_\mathrm{svc}$ with $B_\mathrm{svc}$ bits and $\bm b_\mathrm{cle}$ with $B_\mathrm{cle}$ bits, as shown in \fref{fig:system model}. $\bm b_\mathrm{svc}$ is used for support pattern selection and $\bm b_\mathrm{cle}$ forms the input of the CLE. Hence, $B =
B_\mathrm{svc}+B_\mathrm{cle}$. According to the concept of index modulation (IM)~\cite{cheng2018im}, $B_\mathrm{svc}$ bits generate a unique index $i$ to select a support pattern $\bm \omega_i$ from the CW-pattern set $\bm \Omega = \{\bm \omega_0,\bm
  \omega_1,\cdots,\bm \omega_{I-1}\}$ constituting $I =
2^{B_\mathrm{svc}}$ CW-codewords selected from the constant-weight codebook. The construction of the codebook is introduced in detail in Section \ref{section-ESVC-CWC design}. The pattern $\bm \omega_i \in
\{0,1\}^{N_\mathrm{sv} \times 1}$ satisfies $\|\bm \omega_i\|_0 = w$, representing a binary sparse vector of length $N_\mathrm{sv}$ with $w$ supports. The $B_\mathrm{cle}$ bits of $\bm b_\mathrm{cle}$ are first encoded based on a conventional channel coding scheme, such as Hamming, BCH, RS or polar codes, which support different levels of error correction capability with different trade-off in implementation complexity and/or decoding delay~\cite{rowshan2024channel}. The output of the linear encoder is denoted by $\bm b_\mathrm{s}$. The parameters of these coding schemes in our simulations will be detailed in Section
\ref{section-ESVC-performance and analysis}. Then, the encoded bits in $\bm b_s$ generate a set of $w$ symbols, expressed as $\mathcal S =
\{S_0,S_1,\cdots,S_{w-1}\}$, based on $\mathcal M$-QAM (or PSK) modulation. The symbols in $\mathcal S$ are mapped onto the corresponding supports in $\bm \omega_i$, yielding a sparse vector $\bm v \in \mathbb C^{N_\mathrm{sv}\times 1}$, which contains all information of the transmitted bits in $\bm b$.

\begin{figure}
    \centering
    \includegraphics[width=0.85\linewidth]{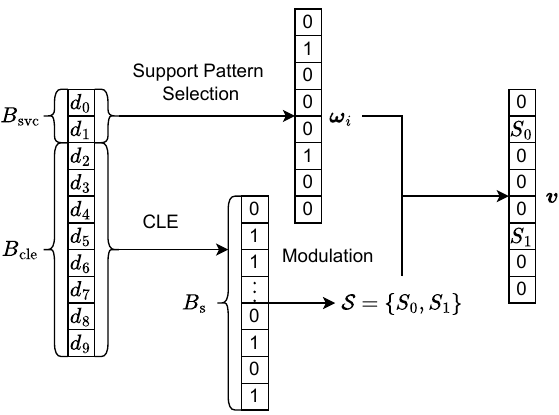}
    \caption{An example for the CW-SVC with $N_\mathrm{sv}=10$, $N=8$ and $w=2$.}
    \label{fig:esvc scheme}
\end{figure}

An example depicted in \fref{fig:esvc scheme} shows the processing of CW-SVC. As shown in \fref{fig:esvc scheme}, $B=10$ bits of data are divided in two sub-blocks: one of $B_\mathrm{svc}=2$ bits on top is used for selecting a pattern $\bm \omega_i$ and the remaining of $B_\mathrm{cle}=8$ bits form the input to the CLE (not drawn in the figure). Assume that a Hamming code is applied. Then, $B_\mathrm{cle}=8$ bits generate $B_s>8$ coded bits. These $B_s$ coded bits are then modulated to form $w=2$ QAM symbols, which are mapped to the two supports in the selected pattern $\bm \omega_i$ to form $\bm v$.

Once the sparse vector $\bm v$ is obtained, as shown in \fref{fig:system model}, it is compressed from length $N_\mathrm{sv}$ to length $N$ of the number of subcarriers, through a Gaussian compressing matrix $\bm C\in \mathbb C^{N\times N_\mathrm{sv}}$, giving
\begin{equation}
    \bm s = \bm{Cv}.
    \label{eq compression}
\end{equation}
The compressed symbols are then input to the A$^2$FDM for further processing. 

Similar to the fast Fourier transform (FFT) in OFDM and discrete affine Fourier transform (DAFT) in AFDM \cite{bemani2021afdm}, discrete augmented affine Fourier transform (DA$^2$FT) \cite{11509294} is the basic building block of A$^2$FDM. As shown in \fref{fig:system model}, the fundamental component of A$^2$FDM, namely the inverse DA$^2$FT (IDA$^2$FT), comprises three components: $\bm \Gamma_\mu$, the inverse FFT (IFFT) matrix $\bm F^H$, and $\bm \Lambda_{c_1}^H$. As previously discussed, the proposed scheme is designed for operation in the scenarios including high mobility and aimed to maintain a low PAPR. To achieve these objectives, in contrast to OFDM, the matrix $\bm \Gamma_\mu$ is constructed based on the subcarrier mapping strategy in the single-carrier frequency division multiple access (SC-FDMA) \cite{yang2009multicarrier} to reduce PAPR. Meanwhile, $\bm
\Lambda_{c_1}^H$ transforms subcarrier waveform from a sinusoidal signal to a chirp signal, causing subcarrier frequency to vary over time. Consequently, this time-varying characteristic enhances the resolution of subcarriers in the Doppler-delay domain, similar to that in AFDM, to achieve diversity gain.

Assuming again that the number of subcarriers is $N$, as shown in \fref{fig:system model}, the IDA$^2$FT is expressed as
\begin{equation}
    \bm A^H = \bm{\Lambda}_{c_1}^H\bm F^H\bm\Gamma_\mu,
    \label{eq Block DAFT}
\end{equation}
where $\bm F^H$ is the IDFT matrix with entries $\left\{\frac{1}{\sqrt{N}}e^{j2\pi mn/N}\right\}$, $\bm
\Lambda_{c_1}^H = \text{diag}\left(e^{j2\pi c_1n^2}\right)$, with $n=0,1,\cdots, (N-1)$ is a diagonal matrix, while $\bm{\Gamma}_\mu$ is a unitary matrix to make $\bm{A}$ a unitary matrix, i.e. satisfying $\bm{A}\bm{A}^H=\bm{A}^H\bm{A}=\bm{I}$, with $\bm{I}$ being an identity matrix.  To mitigate the PAPR of the transmitted signal, $\bm{\Gamma}_\mu$ can be designed as~\cite{11509294}
\begin{align}
  \bm{\Gamma}_\mu =\bm{P\Upsilon}_\mu,
  \label{eq:Gamma}
  \end{align}
by following the principles of SC-FDMA \cite{yang2009multicarrier}. In \eqref{eq:Gamma}, $\bm{\Upsilon}_\mu$ is a $N\times N$ block-diagonal matrix consisting of $\mu$ number of component FFT matrices of $\left(N_\mu \times N_\mu\right)$-dimensional, i.e.
\begin{align}
               \bm{\Upsilon}_\mu  &= \text{diag}\big(\underbrace{ \bm{F}_{N_\mu}, \bm{F}_{N_\mu}, \cdots, \bm{F}_{N_\mu}}_\mu\big)\nonumber\\
            &=\bm I_\mu \otimes \bm F_{N_\mu},
            \label{eq:A2FDM-19}
    \end{align}
where $\otimes$ is Kronecker product, $\bm{F}_{N_\mu}$ is a $\left(N_\mu\times N_\mu\right)$ normalized FFT matrix, $N_\mu =
N/\mu$ is an integer, preferably equalling to $2^i$ for convenience of FFT implementation. In \eqref{eq:Gamma}, $\bm{P}$ is a mapping matrix obeying the interleaved mapping strategy, as that employed in the interleaved FDMA (IFDMA) in SC-FDMA~\cite{yang2009multicarrier}. Hence, we refer to our A$^2$FDM scheme as the interleaved A$^2$FDM (IA$^2$FDM) scheme. Note that the mapping matrix $\bm{P}$ satisfies $\bm{P}^T\bm{P}=\bm{P}\bm{P}^T=\bm{I}$. Therefore, we can readily show that the DA$^2$FT matrix defined in \eqref{eq Block DAFT} is a unitary matrix.

After the $N$-point IDA$^2$FT on $\bm{s}$, the $N$ time-domain samples can be expressed as
\begin{equation}
    x_n = \frac{1}{\sqrt{\mu}}e^{j2\pi c_1n^2} \sum^{\mu-1}_{k=0}s_me^{j2\pi\frac{k n}{N}},~n = 0,1,\cdots,N-1,
    \label{eq xn with IDA2FT}
\end{equation}
where $m = kN_\mu+b$ and $b = n \text{ mod }N_\mu$. The collection of $x_n$ in \eqref{eq xn with IDA2FT} can be written in matrix form as
\begin{equation}
    \bm x = \bm A^H\bm s = \bm \Lambda_{c_1}^H\bm F^H\bm P\bm \Upsilon_\mu\bm s.
    \label{eq IDAFT matrix form}
\end{equation}

As in AFDM, A$^2$FDM introduces a prefix to mitigate the effect of the delay-spread resulting from multipath propagation. Considering the property of chirp waveform, which has the periodicity of \cite{bemani2023affine}
\begin{equation}
    x_{n+uN} = e^{j2\pi c_1(u^2N^2+2uNn)}x_n,\quad u = 0,1,2,\cdots,
\end{equation}
where $u$ is the periodic index, the chirp-periodic prefix (CPP) in A$^2$FDM is given as
\begin{align}
  x_{n} = e^{-j2\pi c_1(N^2+2Nn)}x_{n+N},\quad n = -L_{CPP},\cdots,-1,
    \label{eq cpp}
\end{align}
where $L_{CPP}$ is the length of the CPP.

After the parallel-to-serial (P/S) conversion and adding the CPP, as
shown in \fref{fig:system model}, the signal is transmitted over the
wireless channel, which may be a doubly selective fading channel. The
matrix form of the received signal can be expressed as
\begin{align}
    \bm{y} &= \bm{Hx}+\bm{n}\nonumber\\
    &=\bm{HA}^H\bm{s}+\bm n,
    \label{eq: matrix form received signal}
\end{align}
where $\bm{n} \sim \mathcal{CN}(\bm 0,N_0\bm I)$ is the additive white Gaussian noise (AWGN) vector, and the channel matrix $\bm H$ is given by~\cite{bemani2023affine}
\begin{equation}
    \bm H = \sum^L_{l=1}h_l\bm \Gamma_{CPP_l} \bm \Delta_{f_l} \bm \Pi^{\ell_l},
    \label{eq channel matrix}
\end{equation}
where $L\geq 1$ is the number of propagation paths. In \eqref{eq
  channel matrix}, $h_l$, $f_l$ and $\ell_l$ are the complex gain,
Doppler shift in digital frequency, which is the Doppler shift
normalized by the sampling frequency, and the integer delay,
respectively, associated with the $l$-th path. Following
\cite{bemani2023affine}, the terms in \eqref{eq channel matrix} are
defined as follows: $\bm \Gamma_{CPP_l}$ is an $(N\times N)$ diagonal
matrix given by
\begin{align}
    \bm \Gamma_{CPP_l} = &\mathrm{diag}\left(
    \begin{cases}
        e^{-j2\pi c_1(N^2-2N(\ell_l-n))},& n<\ell_l,\\
        1,&n\geq \ell_l,
    \end{cases},\right.\nonumber\\
     &~~~~~~~~~~~n = 0,1,\cdots,N-1 \bigg),
\end{align}
which for $n<\ell_l$ has the contribution of the phase shifts induced
by the CPP. When $n\geq \ell_l$, the CPP has no effect on the signal,
and hence the corresponding diagonal elements in $\bm \Gamma_{CPP_l}$
are equal to $1$.  $\bm \Delta_{f_l} \triangleq \text{diag}(e^{-j2\pi
  f_ln},n = 0,1,\cdots,N-1)$ accounts for the effect of the Doppler
shifts, while $\bm\Pi^{\ell_l}$ models the effect of the delay of the
$l$-th path, where $\bm \Pi$ is a forward cyclic-shift matrix
expressed as~\cite{bemani2021afdm}
\begin{equation}
    \bm \Pi = 
    \begin{bmatrix}
        0&\cdots&0&1\\
        1&\cdots&0&0\\
        \vdots&\ddots&\ddots&\vdots\\
        0&\cdots&1&0
    \end{bmatrix}_{N\times N}.
\end{equation}

After discarding the CPP, as shown in \fref{fig:system model}, support
and symbol detections are executed to estimate the pattern index $\hat
i$. The total $B = B_\mathrm{svc} + B_\mathrm{cle}$ bits are then
recovered, where $B_\mathrm{svc}$ bits are derived from the index
$\hat i$, and the remaining $B_\mathrm{cle}$ bits are obtained from
the QAM symbols in the support of $\bm \omega_{\hat i}$, after symbol
demodulation and channel decoding.

In the following section, a method for constructing CWC is firstly
presented. Then, the detection method is addressed in Section
\ref{section-ESVC-support detection}.

\section{Design of Constant Weight Sparse Vector Code}\label{section-ESVC-CWC design}

As mentioned in Section \ref{section-ESVC-system model}, the support
patterns are designed based on a CWC. Consider a CWC$(n_w,d_w,w)$,
where $n_w$ denotes the codeword length, $d_w$ is the minimum Hamming
distance between distinct codewords, and $w$ is the weight of the
codewords. In the CWC codebooks designed for the
  proposed system, the codeword weight is constrained by $w\leq
  n_w/2$, and the minimum Hamming distance is deliberately designed to
  increase with the codeword weight. In the proposed CW-SVC/A$^2$FDM
system, CW-codewords are employed to determine the support
patterns. Hence, it follows that $n_w = N_\mathrm{sv}$. Furthermore,
to facilitate detection achieving high reliability, for example in
user activity detection in the grant-free multiple access (GFMA)
scenario, a large $d_w$ is desired. Consequently, the desired
codewords in the CWC are expected to have a relatively large weight in
the CW-SVC with $w\leq N_\mathrm{sv}/2$. This illustrates the
fundamental difference between the proposed CW-SVC scheme and the
other existing ESVC schemes
\cite{kim2020enhanced,11409473,11432793}. Furthermore, in the existing
ESVC schemes, the number of the candidate support patterns is from
$\binom{N_\mathrm{sv}}{w}$ with the minimum Hamming distance as low as
$2$, whereas the number of codewords of CWC$(n_w,d_w,w)$ is
upper-bounded by the Johnson bound \cite{1057714}. Specifically, for
$n_w\geq w>0$, the Johnson bound is given by
\begin{equation}
A(n_w,d_w,w) \leq\left\lfloor \frac {n_w}{w}A(n_w-1,d_w,w-1)\right\rfloor,
\label{eq johnson bound}
\end{equation}
where $A(n_w,d_w,w)$ denotes the maximum possible number of codewords of CWC$(n_w,d_w,w)$. Therefore, the number of
candidates in a CWC-assisted support pattern set for our
CW-SVC/A$^2$FDM system is significantly smaller than $\binom{N_\mathrm{sv}}{w}$, in particular when $N_\mathrm{sv}$ and $w$ are relatively big. 

Regarding the construction of constant-weight codebooks, when $d_w=2w$ holds, the CWC can be constructed directly. In this case, if $N_\mathrm{sv}$ is divisible by $w$, the $N_\mathrm{sv}$ positions are partitioned into $N_\mathrm{sv}/w$ disjoint support sets of size $w$, with each set serving as one codeword, so that any two codewords have disjoint supports and the minimum Hamming distance is exactly $d_w=2w$. However, when this property is no longer satisfied, i.e. for $d_w<2w$, in this paper the CWC is generated by the ``from dimension to weight (FDTW)'' construction \cite{6979861}.
With the FDTW
method, a constant dimension code (CDC) is first generated and then
the codewords of the CWC are obtained from its cosets of each subspace. In detail, let us
consider a vector space $\mathbb F^{n_d}_q$ with $q^{n_d}$ vectors,
where $n_d$ is the dimension of $\mathbb F^{n_d}_q$, i.e. the length
of vectors in $\mathbb F^{n_d}_q$. The elements in these vectors
belong to $\{0,1,\ldots,q-1\}$. According to \cite{6979861}, given a
non-negative integer $k\leq n_d$, the set of all subspaces of $\mathbb
F^{n_d}_q$ with dimension $k$ is known as a Grassmannian, which can be
represented by $\mathcal G_q(n_d, k)$. The distance between such two
subspaces $\mathcal U$ and $\mathcal V$ is given by
\cite{koetter2008coding,silva2008rank,etzion2011error}
\begin{equation}
    d(\mathcal U,\mathcal V) \triangleq \text{dim }\mathcal U + \text{dim }\mathcal V -2\text{dim }(\mathcal U\cap \mathcal V)=2t,
\end{equation}
for all $\mathcal U,\mathcal V \in \mathcal G_q(n_d, k)$. Denote
$\mathcal C \subseteq \mathcal G_q(n_d, k)$. If there exists a minimum
dimension distance $d_d$, allowing $d(\mathcal U,\mathcal V)\geq d_d$
for all $\mathcal U,\mathcal V\in \mathcal C$, we say that $\mathcal
C$ is a $(n_d,d_d,k)$ CDC set in $\mathcal G_q(n_d, k)$ and $|\mathcal
C|$ is the size of this set, i.e. the number of subspaces.

Let $\mathcal X$ be a subspace of $\mathcal C$ with dimension $k$,
which contains $q^k$ vectors of $n_d$-dimensional. Then, the cosets of
$\mathcal X$ are given by \cite{6979861}
\begin{equation}
    \bm\beta+\mathcal X = \{\bm\beta+\bm\gamma: \bm\gamma\in \mathcal X\}.
    \label{eq cosets}
\end{equation}
{Note that in \eqref{eq cosets}, although $\bm \beta$ can be chosen
  from the whole space $\mathbb{F}_q^{n_d}$, different choices of $\bm
  \beta$ may generate the same coset. In fact, since
  $|\mathbb{F}_q^{n_d}|=q^{n_d}$ and $|\mathcal{X}|=q^k$, there are
  only $\frac{|\mathbb{F}_q^{n_d}|}{|\mathcal{X}|}
  =\frac{q^{n_d}}{q^k}=q^{n_d-k}$ distinct cosets of $\mathcal{X}$,
  which partition $\mathbb{F}_q^{n_d}$. Note furthermore that the
  principal distinction between a subspace and its cosets is that a
  subspace always contains the zero vector, while the cosets do not.}

Once the cosets of a subspace are obtained, their vectors are mapped
to decimal numbers, which specify the indices of \enquote{$1$} in a
$q^{n_d}$ length vector, and all other positions are filled with
\enquote{$0$}, giving the supports of vectors. Consequently, each
coset corresponds to a codeword of length $n_w=q^{n_d}$ and weight
$w=q^k$, i.e. a sparse vector of length $n_w$ and sparsity
$w$. By taking the union of the cosets of all
  subspaces in $\mathcal C$, a CWC with $I_\mathrm{fdtw}=|\mathcal
  C|q^{n_d-k}$ codewords is obtained. Note that, in the proposed
  CW-SVC system, only $I = 2^{B_\mathrm{svc}} \leq I_\mathrm{fdtw}$
  codewords are utilized as the candidate support patterns, where
  $B_\mathrm{svc} = \lfloor \log_2 I_\mathrm{fdtw}\rfloor$. The
minimum Hamming distance of this CWC is given by \cite{6979861}
\begin{equation}
    d_w = 2\left(q^k-q^{k-t}\right),
    \label{eq fdtw hamming distance}
\end{equation}
where $t\in [0,k]$ to ensure $d_w$ remains non-negative. The
parameters of a general CWC constructed by the FDTW method is shown in
\tref{tab:parameters CDC-CWC}.
\begin{table}
    \centering
    \caption{Parameter relationships between CDC and the CWC
      constructed from FDTW method.}
    \begin{tabular}{|c|c|}\hline
         CWC codeword length& $n_w = q^{n_d}$\\\hline
         CWC codeword weight& $w = q^k$\\\hline
         CWC's minimum Hamming distance& $d_w = 2(q^k-q^{k-t})$\\ \hline
 CWC  size&$I_\mathrm{fdtw} = |\mathcal C|q^{n_d-k}$\\\hline
    \end{tabular}
    \label{tab:parameters CDC-CWC}
\end{table}

To further illustrate the above CWC construction principle, an example is provided in the following. Note that the values of $n_w$ and $w$ are usually
determined by the requirements of the sparse vectors in the proposed
CW-SVC/A$^2$FDM system.  When these values are given, the minimum
Hamming distance should be as large as possible to achieve a higher
reliability, provided that the number of codewords in CWC is sufficient
for supporting user and/or data rate requirements. Therefore, with
this example, we assume that a CWC$(16,6,4)$ is needed, where $n_w =
16$, $d_w = 6$ and $w = 4$. Hence, consider a vector space
$\mathbb{F}_2^4$, where $q=2$ and $n_d=4$, which leads to $n_w = 2^4 = 16$,
$k$ can be chosen to be $k=2$, which results in $w = 2^k = 4$. Then, a CDC
$\mathcal{C}\subseteq G_2(4,2)$ is chosen for the FDTW
construction. According to \tref{tab:parameters CDC-CWC}, $d_w =
2\left(2^2-2^{(2-t)}\right)=6$, giving $t=2$. Thus, the minimum
subspace distance $d_d=2t = 4$. For simplicity, in this example, we
select two subspaces $\mathcal{X}_1,\mathcal{X}_2\in\mathcal{C}$ of
two-dimensional, generated over $\mathbb F_2$ by two linearly
independent vectors, given by
\begin{align}
    \mathcal X_1 &= \mathrm{span}_{\mathbb F_2}\{(0001,0010)\}\nonumber\\
    &=\{a_1(0001)+a_2(0010):a_1,a_2\in\mathbb \{0,1\}\} \nonumber\\
    &=  \{0000,0001,0010,0011\}\nonumber\\
    \mathcal X_2 &= \mathrm{span}_{\mathbb F_2}\{(0100,1000)\} \nonumber\\
    &=\{a_1(0100)+a_2(1000):a_1,a_2\in\mathbb \{0,1\}\} \nonumber\\ 
    &= \{0000,0100,1000,1100\},
    \label{eq subspace}
\end{align}
each of which contains $q^k=2^2=4$ vectors of length $n_d = 4$. As
seen in \eqref{eq subspace}, $\mathcal X_1$ and $\mathcal X_2$ share
only the all-zero vector. Hence, $\operatorname{dim}(\mathcal
X_1\cap\mathcal X_2) = 0$. Therefore, the subspace distance between
$\mathcal X_1$ and $\mathcal X_2$ is
\begin{align}
    d(\mathcal X_1,\mathcal X_2) &= \operatorname{dim}(\mathcal X_1)+\operatorname{dim}(\mathcal X_2)-2\operatorname{dim}(\mathcal X_1\cap\mathcal X_2)\nonumber\\&= 2+2-2\times 0=4.
\end{align}

As shown in \eqref{eq subspace}, \enquote{$1$} in $\mathcal X_1$
occurs at the third and fourth indices of the $4$-length vectors in
$\mathcal X_1$, indicating the activation of these indices.
{Therefore, for $\mathcal{X}_1$, the convenient choices of distinct
  coset representatives are obtained by varying the inactive indices, i.e. the first and second indices that are always zero in the $4$-length vectors in $\mathcal{X}_1$.}  Note that the zero vector
is always considered as a choice of $\bm\beta$, which returns the
subspace itself. Hence, for $\mathcal X_1$, a set of $q^{n_d-k}=2^{4-2}=4$ distinct coset representatives can be chosen as $\{0000,1000,0100,1100\}$, which results in
\begin{align}
    &(0000)+\mathcal X_1 = \{0000,0001,0010,0011\}\nonumber\\
    &(1000)+\mathcal X_1 = \{1000,1001,1010,1011\}\nonumber\\
    &(0100)+\mathcal X_1 = \{0100,0101,0110,0111\}\nonumber\\
    &(1100)+\mathcal X_1 = \{1100,1101,1110,1111\}.
    \label{eq cosets of X1}
\end{align}

Similarly, for $\mathcal X_2$, {the 4 distinct coset representatives
  are} $\{0000,0010,0001,0011\}$ and the corresponding cosets of
$\mathcal X_2$ are
\begin{align}
    &(0000)+\mathcal X_2 = \{0000,0100,1000,1100\}\nonumber\\
    &(0010)+\mathcal X_2 = \{0010,0110,1010,1110\}\nonumber\\
    &(0001)+\mathcal X_2 = \{0001,0101,1001,1101\}\nonumber\\
    &(0011)+\mathcal X_2 = \{0011,0111,1011,1111\}.
    \label{eq cosets of X2}
\end{align}

Then, the $4$-length vector elements in these cosets are converted
from binary to decimal to represent the support indices in the CW-codewords,
giving a support pattern set containing the $8$ codewords of the
CWC(16,6,4).  \tref{tab:support patterns} lists the support indices of
the CW-codewords generated based on the cosets in \eqref{eq cosets of X1} and
\eqref{eq cosets of X2}. For example, the four vector elements in
$(0000)+\mathcal X_1$ are mapped to $\{0,1,2,3\}$, the four vector
elements in $(0011)+\mathcal X_2$ are mapped to
$\{3,7,11,15\}$. Similarly, the mapping results in the other cases can
be obtained.  According to \eqref{eq fdtw hamming distance}, the
minimum Hamming distance of this CWC with eight codewords of length
$n_w=16$ and $w=4$ is $d_w = 6$.

\begin{table}
    \centering
    \caption{Support indices of the CWC(16,6,4) constructed from FDTW
      with cosets \eqref{eq cosets of X1} and \eqref{eq cosets of
        X2}.}
    \begin{tabular}{|c|c|}\hline
 support pattern&support indices\\\hline
         $\bm \omega_0$&  $\{0,1,2,3\}$  \\\hline
         $\bm \omega_1$&  $\{8,9,10,11\}$\\\hline  
         $\bm \omega_2$&  $\{4,5,6,7\}$\\\hline  
         $\bm \omega_3$&  $\{12,13,14,15\}$\\\hline  
         $\bm \omega_4$& $\{0,4,8,12\}$\\\hline
         $\bm \omega_5$& $\{2,6,10,14\}$\\\hline
         $\bm \omega_6$& $\{1,5,9,13\}$\\\hline
         $\bm \omega_7$& $\{3,7,11,15\}$\\\hline
    \end{tabular}
    
    \label{tab:support patterns}
\end{table}

As shown in \tref{tab:parameters CDC-CWC} and \tref{tab:support patterns}, the number of codewords in the CWC constructed from FDTW is limited, due
to the introduced redundancy for reliability protection. Therefore, a
detection algorithm with relatively high complexity can be employed
for the support detection with low latency. Afterwards, the QAM symbol
detection can be implemented using the principle of, for example the
minimum mean square error (MMSE) estimation. The detailed detection
algorithms will be discussed in the following section.

\section{Support and QAM Symbol Detection}\label{section-ESVC-support detection}

In this section, we first present the proposed detection algorithm for
joint support-pattern identification and QAM symbol detection in the
CW-SVC/A$^2$FDM system. Then, we analyze the detection scheme's
computational complexity and processing latency, which are compared
with that of the conventional orthogonal
matching pursuit (OMP) based detection algorithms employed
by the ESVC schemes.

\subsection{Detection Algorithm}\label{subsection-ESVC-support detection-support detection}

It has been well-recognized that URLLC relies on relatively short
packet transmission due to the strict limit on transmission latency,
but requiring ultra high reliability, making it extremely challenging
in practical implementation. In some of the existing ESVC schemes
\cite{zhang2025block,kim2020enhanced,ji2018sparse,sinha2024generalized},
message transmission and detection are based on the CS
principles~\cite{baraniuk2007compressive,qaisar2013compressive,foucart2013invitation,choi2017compressed}. To
guarantee the required reliability of message recovery, in CS-based
detection, different pattern candidates must be sufficiently
distinguishable. This requires that the mutual correlation between the
columns of CS matrix $\bm C$ is sufficiently low and the sparsity $w$
is small and limited. Furthermore, due to the strict reliability
requirement of URLLC, the modulation schemes of high order, such as 64
QAM or above, may not be acceptable. This is because to transmit the
same amount of information, while using a higher order modulation
scheme increases transmission rate and hence reduces latency, but it
results in a higher error rate, which penalizes the reliability of
URLLC. To mitigate this dilemma, conventional ESVC schemes generally
employ the modulation (typically QAM) of relatively low-order to
maintain reliable QAM detection, meaning that only a small number of
data bits are conveyed by QAM. Accordingly, for a given information
payload and a sparsity level, a relatively big number of bits must be
conveyed by the support patterns.  Since $I=2^{B_\mathrm{svc}}$
candidate support patterns are required to convey $B_\mathrm{svc}$ bits per symbol, the complexity of sparse recovery increases exponentially with
$B_\mathrm{svc}$.  Additionally, the modulation of QAM symbols on the
supports of sparse vectors further enlarge the search space of signal
detection. In this case, the employment of maximum likelihood (ML)
detection that achieves optimal performance is impractical due to its
extremely high complexity. Instead, the support detection is typically
achieved by the suboptimal greedy CS algorithms, such as OMP \cite{tropp2007signal} or multipath matching
pursuit (MMP) \cite{kwon2014multipath}, which, in comparison with ML
detection, sacrifices some reliability performance while significantly
reducing the detection complexity.

In contrast to the conventional ESVC schemes, the support patterns in
the proposed CW-SVC/A$^2$FDM system are determined by CW-codewords
that are well-regularized. Specifically for URLLC applications, the
CW-codewords with large minimum Hamming distances are expected, which
result in limited number of candidates, allowing to employ the
detection schemes of relatively high-complexity in
practice. Furthermore, when the Hamming distance of a CWC increases,
its weight $w$ also increases, as demonstrated in
Section~\ref{section-ESVC-CWC design}.  Hence, compared to the
conventional ESVC assisted schemes that rely on a small number of
supports to convey high-order QAM symbols, using our CW-SVC scheme most of the data can be conveyed by the relatively low-order QAM symbols,
with only a small portion of the data will be transmitted by the
candidate support patterns. Therefore, in this paper, we propose to
employ the detection schemes of a relatively high complexity but low
latency. Specifically, we propose a candidate-wise MMSE-assisted
residual detection. With this detection, the receiver first generates
an effective sensing matrix for each of the candidate support
patterns, from which an MMSE estimator is then obtained for detecting
the corresponding QAM symbols. Subsequently, the received signal is
reconstructed for each of the candidate support patterns, and then the residual between the reconstructed and actual received signals is
calculated. Consequently, by comparing the residuals corresponding to
all candidate support patterns, the receiver determines the
transmitted support pattern's index $\hat i$ that corresponds to the
support pattern yielding the minimum residual. After the support
pattern $\bm \omega_{\hat{i}}$ is detected, the QAM symbols conveyed
on the supports can be extracted for subsequent channel decoding. From
the above description we can know that the detection for different
candidate support patterns can be implemented in parallel to reduce
the receiver processing latency.

To formulate the proposed detection, upon combining \eqref{eq compression} and \eqref{eq:
  matrix form received signal}, and assuming that the $i$-th pattern
(CW-codeword $\bm \omega_i$) is actually selected, the received signal in the
CW-SVC/A$^2$FDM  system can be expressed as
\begin{align}
    \bm y &=\bm{HA}^H\bm s+\bm n\nonumber\\
    &= \bm{HA}^H\bm{Cv}+\bm n\nonumber\\
    &= \bm{HA}^H\bm C_i\bm v_i+\bm n,
    \label{eq received signal jd}
\end{align}
where $\bm v_i$ is the subvector of the sparse vector $\bm v$ restricted by the support positions determined by $\bm\omega_j$, as seen in
Table~\ref{tab:support patterns} for example, where each support
conveys one QAM symbol, while $\bm C_i$ is the submatrix of $\bm C$,
consisting of the columns indexed by the elements in $\bm
\omega_i$. 

However, the receiver does not know that the
  transmitter selected the $i$-th pattern. Therefore, in order to
  fulfill the detection, the receiver first assumes that the activated
  pattern is one of $\bm\Omega=\{\bm \omega_j,~ \forall
  j=0,1,\cdots,I-1\}$. Then, considering that the receiver has the
  full knowledge of $\bm A$ and $\bm H$, it can implement the
  MMSE-assisted processing to mitigate the interference between
  different supports. Specifically, the MMSE estimation matrix
  corresponding to the assumed pattern $\bm \omega_j$ can be
  calculated as
\begin{equation}
\bm W_j=\bm G^H_j\left(\bm G_j\bm G_j^H+\frac{1}{\gamma}\bm{I}_N\right)^{-1},\quad j = 0,1,\cdots,I-1,
\label{eq equalization matrix-jd}
 \end{equation}
where $\bm G_j = \bm{HA}^H\bm C_j$ is the effective sensing matrix. In
\eqref{eq equalization matrix-jd}, $\gamma$ is the signal-to-noise
ratio (SNR) and $\bm I_N$ is an identity matrix of size $(N\times N)$.
Consequently, when the receiver assumes that $\bm{v}_j$ was
transmitted (although the actually transmitted vector was $\bm{v}_i$,
necessitating the clarification stage shown in \eqref{eq score}), the
QAM symbols corresponding to the support of $\bm\omega_j$ can be
estimated as
 \begin{equation}
     \tilde{\bm S}_j = \bm W_j\bm y,\quad j=0,1,\cdots,I-1.
     \label{eq equalization-jd}
 \end{equation}

To detect the QAM symbols, the bias from the MMSE processing is
corrected by normalizing $\tilde{\bm S}_j$ using $\bm
\Gamma_j=\textrm{diag}\left(\bm W_j\bm G_j\right)$,
giving
\begin{equation}
\tilde{\bm S}_j^\mathrm{norm}=\bm \Gamma_j^{-1}\tilde{\bm S}_j.
\label{eq normalization-jd}
\end{equation}

Then, for each $\bm \omega_j$, a penalty is computed as the sum of the
absolute errors between the received signal $\bm y$ and the
reconstructed signal $\bm G_j\tilde{\bm
  S}_j^\mathrm{norm}$. Consequently, the index of the selected sparse
vector (CW-codeword) by the transmitter can be determined as
\begin{equation}
    \hat i = \operatorname*{arg\,min}_{j=0,1,\ldots,I-1} \left\|\bm y-\bm G_j\tilde{\bm S}_j^\mathrm{norm}\right\|^2.
    \label{eq score}
\end{equation}
Accordingly, the data bits conveyed by the index $\hat i$ of the
sparse vector can be recovered as $\hat {\bm b}_\mathrm{svc}$.

Once the index $\hat i$ of the activated sparse vector is detected,
the data bits $\hat{\bm b}_\mathrm{s}$ conveyed by the QAM symbols
conveyed on the supports of $\omega_{\hat{i}}$ can be demodulated from
$\tilde{\bm S}_{\hat{i}}^\mathrm{norm}$. Afterwards, the bits of
$\hat{\bm b}_\mathrm{cle}$ can be recovered from the corresponding
linear decoding of $\hat{\bm b}_\mathrm{s}$. In particular, if polar
coding is employed as the CLE, the inputs of the decoder need to be
the log-likelihood ratio (LLR) values of the detected symbols, which
can be expressed as
\begin{equation}
L_u=\gamma\left[\min_{a\in\mathcal A_q^{(1)}}\left|\tilde{S}_{\hat i,k}^{\mathrm{norm}}-a \right|^2-\min_{a\in\mathcal A_q^{(0)}}\left|\tilde{S}_{\hat i,k}^{\mathrm{norm}}-a\right|^2\right],
\label{eq:llr-jd}
\end{equation}
with $u = k\log_2(\mathcal M)+q$, where $\tilde{S}_{\hat i,k}^{\mathrm{norm}}$ is the $k$th symbol in $\tilde{\bm S}_{\hat{i}}^{\mathrm{norm}}$, $k = 0,1,\cdots,w-1$, and $a$ is one of the possible constellation point of the $\mathcal M$-QAM
modulation employed, where the $q$th bit of a QAM symbol is either
\enquote{0} or \enquote{1} with $q = 1,\cdots,\log_2(\mathcal M)$, and
$\mathcal A_q^{(\cdot)}$ denotes the set of constellation
points.

Overall, the detection algorithm is summarised in \alref{alg:jd}. Note that, when applied to URLLC, the detection in terms of $i =
0,\cdots,I-1$ can be implemented in parallel to reduce latency.

\begin{algorithm}
\caption{Joint support and symbol detection.}
\label{alg:jd}
\begin{algorithmic}[1]
\REQUIRE $\bm y$,  $\bm H$, $\bm A$, $\bm C$, $\bm \Omega$,$\gamma$
\ENSURE $\hat{i}$, $\hat{\bm b}_\mathrm{svc}$, $\hat{\bm b}_\mathrm{cle}$
\FOR{$j = 0,\cdots,I-1$}
    \STATE Form $\bm C_j$ using the columns of $\bm C$ corresponding to the supports in $\bm \omega_j$,
    \STATE Form the effective sensing matrix $\bm G_j=\bm{HA}^H\bm C_j$,
    \STATE Compute the MMSE estimator of \eqref{eq equalization matrix-jd} and estimate symbols as \eqref{eq equalization-jd},
    \STATE Correct the bias as \eqref{eq normalization-jd},
    \STATE Calculate the penalty
    \[
    \mathrm{penalty}(j) = \left\|\bm y-\bm G_j\tilde{\bm S}^{\mathrm{norm}}_j\right\|^2
    \]
\ENDFOR
\STATE Determine the best CW-SVC index as:
\[
\hat{i}=\operatorname*{arg\min}_j ~\mathrm{penalty}(j)
\]
\STATE Recover $\hat{\bm b}_\mathrm{svc}$ from $\hat i$
\STATE Recover $\hat{\bm b}_\mathrm s$ from $\tilde{\bm S}^{\mathrm{norm}}_{\hat i}$
\STATE Recover $\hat{\bm b}_\mathrm{cle}$ by linearly  decoding $\hat{\bm b}_\mathrm{s}$
\end{algorithmic}
\end{algorithm}

\subsection{Complexity and Latency Analysis}\label{subsection-ESVC-support detection-complexity analysis}

In the conventional SVC/ESVC schemes, usually the OMP is implemented to detect the active supports based on the whole sensing dictionary $\bm
G\in \mathbb{C}^{N\times N_\mathrm{sv}}$ \cite{zhang2023uniquely}. In each OMP iteration, the
dominant operation is given by the correlation between the residual
and all the $N_\mathrm{sv}$ columns in $\bm G$, which requires
$\mathcal O(NN_\mathrm{sv})$ operations. In addition, a least-square
aided update is performed over the identified support set, whose
complexity is approximately $\mathcal{O}(Nk^2+k^3)$ at the $k$-th
iteration. Therefore, after $w$ iterations, the overall complexity of
the OMP can be approximated as
\begin{equation}\label{eq:complexity-OMP}
    \mathcal O_\mathrm{OMP} = \mathcal O(wNN_\mathrm{sv}+Nw^3+w^4).
\end{equation}

In contrast, in the proposed detection scheme for CW-SVC/A$^2$FDM  systems, the
support patterns are restricted to a fixed CWC codebook with $I$
candidate codewords. For each candidate $\bm \omega_j$, the
corresponding effective sensing matrix $\bm G_j$ is formed, and a QAM
symbol sequence is estimated by the MMSE estimator of \eqref{eq equalization matrix-jd}.
Since the matrix inversion in the MMSE estimator is performed for each
candidate, it requires $\mathcal O(N^3)$ operations per candidate, while the
subsequent symbol estimation and penalty calculation require approximately
$\mathcal O(Nw)$ operations per candidate. Therefore, the overall complexity of
the proposed detector is dominated by the per-candidate matrix inversions and is
\begin{equation}\label{eq:complexity-CW-SVC}
    \mathcal O_\mathrm{proposed} = \mathcal O(IN^3),
\end{equation}
which grows linearly with the number of candidates $I$. When comparing
\eqref{eq:complexity-OMP} with \eqref{eq:complexity-CW-SVC}, for a small
$I$, the complexity of the proposed detector is manageable, and the candidate-wise
operations can be implemented in parallel to reduce the processing latency. For a
large $I$, e.g. $I=2048$, the detection complexity becomes high, which is
consistent with the serial-detection results discussed in the performance
section.


In addition to ultra-high reliability, low latency is another key
requirement of URLLC. In a communication system, the end-to-end
latency consists of several key component delays, including queuing
delay, scheduling/access delay, transmitter-side processing delay,
over-the-air transmission delay, propagation delay, receiver-side
processing delay, and feedback/retransmission delay~\cite{8472907}. In
the proposed scheme, the receiver-side processing latency can be
significantly reduced in comparison with the OMP-based
method. Specifically, in OMP-based scheme and most other sparse
recovery algorithms, the received signal has to be processed through
multiple iterations to achieve detection. In each iteration, the
residual is updated and compared with all candidate codewords or atoms
to determine the final detected signal. Since the computational
results in the previous iteration is required for the subsequent
iterations, this procedure must be performed in a sequential
manner. Furthermore, while a larger number of iterations can improve
detection performance, it also leads to the increase of receiver-side
processing latency, which increases linearly with the number of
iterations.  In contrast, our proposed detection
  algorithm does not rely on iterative operations. The detection
  metrics associated with different candidates are independent of each
  other, and hence support both serial and parallel implementations,
  providing flexibility for different application
  scenarios. Specifically, when hardware resources are constrained
  while the latency requirement is relaxed, the candidate metrics can
  be evaluated in serial. In contrast, when the latency requirement is
  stringent while hardware resources are not the limiting factor, as
  in URLLC scenarios, the candidate metrics can be evaluated in
  parallel to meet the latency requirement.

\section{Performance Results and Analysis}\label{section-ESVC-performance and analysis}
In this section, we present the simulation results to demonstrate the
performance of our proposed CW-SVC/A$^2$FDM  system. Unless
specified, in our simulations, the length of sparse vector is
$N_\mathrm{sv} = 128$, the number of subcarriers is $N=64$ and there
are $B=34$ information bits transmitted per symbol, where
$B_\mathrm{svc}=2$ for CW-codewords selection and $B_\mathrm{cle}=32$ for
classical linear encoding and further QAM/PSK modulation. Note that in
the case of using no channel coding, namely uncoded CW-SVC,
$B_\mathrm{s}=32$ is implemented for the same bit throughput. The CW-codewords
used in simulations are generated by the FDTW construction detailed in
Section \ref{section-ESVC-CWC design} and an example is given in
\apref{Appendix-CWC}. The delays of doubly-selective fading channels
normalized by the sampling duration are randomly distributed in the
integer set $[1,\ell_{\max}=30]$, where $\ell_{\max}$ denotes the
maximum normalized delay-spread. The Doppler shifts normalized by
subcarrier spacing $\Delta f$ are distributed in
$[-\nu_{\max},\nu_{\max}]$, set according to
$\nu_l=\nu_{\max}\text{cos}(\varphi_l)$ with $\varphi_l$ uniformly
distributed in $[-\pi, \pi]$. The value of $c_1$ is set as $c_1 =
\frac{2\nu_{\max}+1}{2N}$ to achieve the maximum diversity order.
Unless specified otherwise, $\mu$ is set to 4 for the generation of $\bm \Gamma_\mu$.

\begin{table}[t]
    \centering
\caption{System parameters considered
 in \fref{fig: scheme comparison}, \fref{fig: error source} and \fref{fig: mu comparison}.}
\label{tab:scheme comparison}
    \begin{tabular}{|c|c|c|c|c|}\hline
         & $B_\mathrm{svc}$ & $B_\mathrm{s}/B_\mathrm{cle}$ &weight&Modulation\\\hline
         \makecell{Conventional\\ ESVC}& $B_\mathrm{svc} =24$& $B_\mathrm{s} =10$ &$w=5$&4 QAM\\\hline
         \makecell{Uncoded\\ CW-SVC}& $B_\mathrm{svc} =2$& $B_\mathrm{s} =32$ &$w=32$&BPSK\\ \hline
 \makecell{polar coded\\ CW-SVC}&$B_\mathrm{svc} =2$&\makecell{$B_\mathrm{cle} =32$\\$B_\mathrm{s} =64$} &$w=32$&4 QAM\\\hline
    \end{tabular}
\end{table}

\begin{figure}
    \centering
    \includegraphics[width=0.85\linewidth]{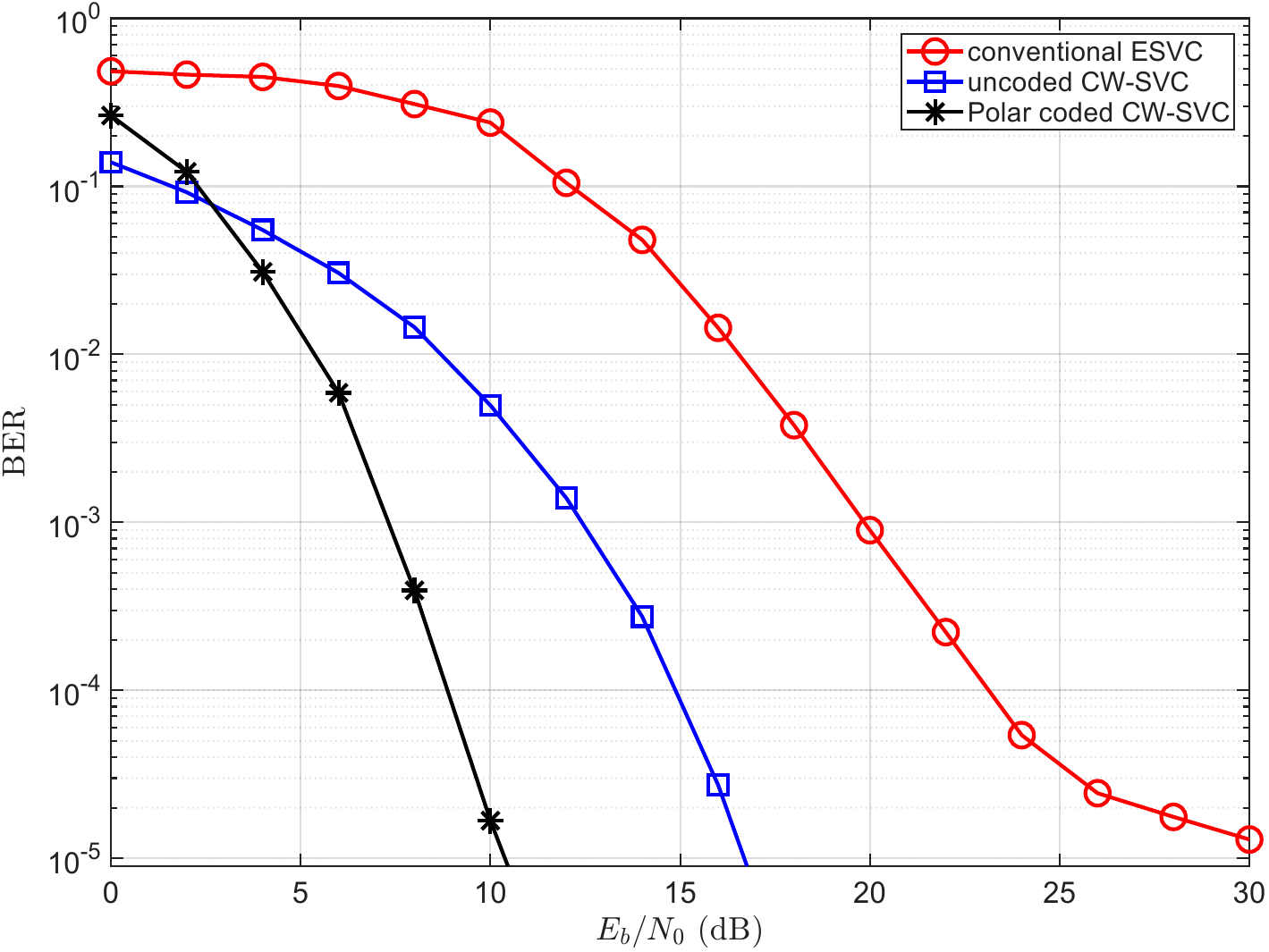}
    \caption{BER performance comparison between the conventional ESVC,
      uncoded CW-SVC and the polar coded CW-SVC schemes operated with A$^2$FDM.}
    \label{fig: scheme comparison}
\end{figure}

\fref{fig: scheme comparison} demonstrates the BER performance of the
A$^2$FDM systems aided by the conventional ESVC and the proposed
CW-SVC without or with the polar CLE. In this comparison, the
CWC$(128,64,32)$ is employed and the number of CW-codewords is $I=4$. The
detailed parameters are listed in \tref{tab:scheme comparison}. The
conventional ESVC uses OMP algorithm for sparse recovery. As shown in
\fref{fig: scheme comparison}, the BER of the conventional ESVC scheme
exhibits an error floor at about the BER of $10^{-5}$, which is due to
the employment of the greedy algorithm. In contrast, with a similar
data throughput and channel state information (CSI), the proposed
CW-SVC scheme without or with polar CLE outperforms the conventional
ESVC.  Specifically, the CW-SVC without CLE enables a BER of $10^{-5}$
at about 16.5 dB, which is more than 14~dB lower than that by the
conventional ESVC scheme. Furthermore, when the CW-SVC is enhanced by
polar CLE, about a further 6~dB of SNR gain can be obtained, achieving
a BER of $10^{-5}$ at about 10.5 dB.

\begin{figure}
    \centering
    \includegraphics[width=0.85\linewidth]{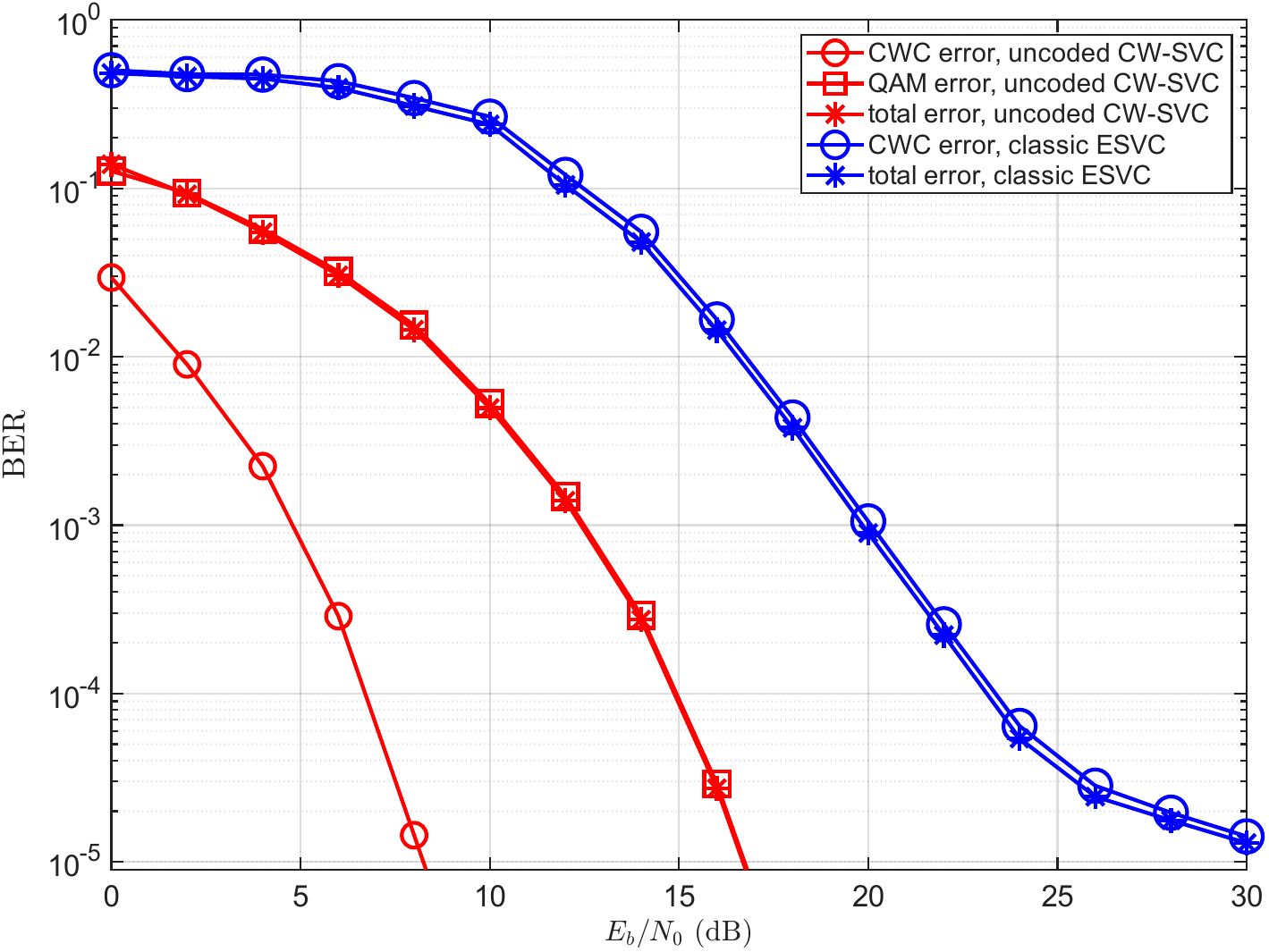}
    \caption{BER of SVC, QAM and overall of the conventional ESVC and
      the uncoded CW-SVC.}
    \label{fig: error source}
\end{figure}

In \fref{fig: error source}, we analyze the BER performance for the SVC, QAM
and overall, when the conventional ESVC scheme and the proposed CW-SVC scheme
without CLE are compared. The legends in \fref{fig: error source} are
defined as follow:
\begin{itemize}
    \item SVC error: errors generated solely in the support pattern
      detection;
    \item QAM error: errors generated solely in the QAM demodulation
      after the support patterns are correctly detected;
    \item Overall error: include the errors in both the above cases.
\end{itemize}
As shown in \fref{fig: error source}, for the proposed CW-SVC scheme,
the BER of SVC is significantly lower than the BER of QAM, while the
overall BER is similar as that of QAM. This explains that in the
proposed CW-SVC scheme, the overall BER performance is dominated by
the errors generated from QAM demodulation. Therefore, in the proposed
CW-SVC/A$^2$FDM systems, appropriate CLE is required to protect the
information conveyed by QAM, to reduce the overall BER.  In contrast,
for the conventional ESVC scheme, the BER of SVC is close to the
overall BER, meaning that the overall system's BER is dominated by the
SVC errors. The results shown in \fref{fig: error source} are
predictable, because in the proposed CW-SVC/A$^2$FDM system, the
support patterns of sparse vectors are protected by the CWC. However,
in the conventional ESVC assisted systems, the support patterns of
sparse vectors are not protected, contributing the majority of errors.

\begin{figure}
    \centering
    \includegraphics[width=0.85\linewidth]{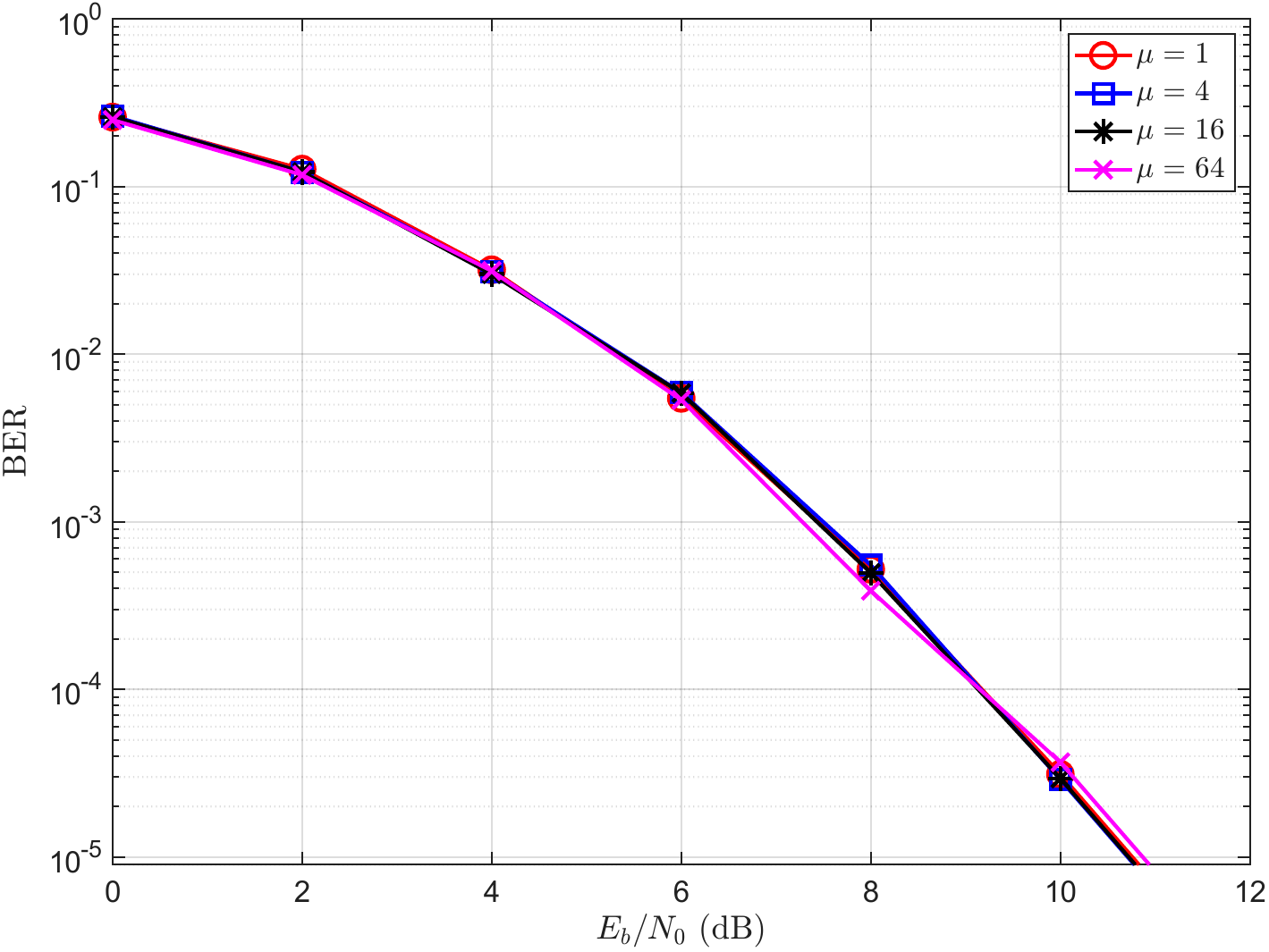}
    \caption{BER performance comparison of the CW-SVC/A$^2$FDM
      systems, when $\mu=1,4,16$ and $64$ are respectively applied for
      the $\bm{\Gamma}_\mu$ matrix in A$^2$FDM.}
    \label{fig: mu comparison}
\end{figure}

In \fref{fig: mu comparison}, the BER performance of the
CW-SVC/A$^2$FDM system with $\mu = 1,4,16$ and $64$ applied in the
$\bm{\Gamma}_{\mu}$ matrix for A$^2$FDM is compared. Clearly, there is
not much difference in these results. According to the principles
analyzed in \cite{11509294}, with the number of subcarriers $N=64$,
when $\mu=1$, the A$^2$FDM is reduced to a single-carrier
system. Then, the A$^2$FDM becomes a multicarrier system from $\mu=2$
and the number of distinguishable subcarriers increases with the
growth of $\mu$, until the system becomes the AFDM when $\mu=64$. For
all the values of $\mu$, as long as the value of $c_1$ is
appropriately selected to achieve the maximum diversity gain, the
effect of $\mu$ is little on the BER performance of A$^2$FDM. However,
we should note that the PAPR of A$^2$FDM signals increases with the
increase of the value of $\mu$, i.e. no PAPR at $\mu=1$, while highest
PAPR at $\mu=64$, which is the PAPR of AFDM
signals. Moreover, the low-PAPR property of
  A$^{2}$FDM due to its embedded SC-FDMA-based IDA$^{2}$FT was studied
  in \cite{11509294}. Since the CW-SVC-generated signal is processed
  by the same A$^{2}$FDM modulator, the employment of CW-SVC does not
  alter the low-PAPR property of the A$^{2}$FDM signalling.

\begin{figure}
    \centering
    \includegraphics[width=0.85\linewidth]{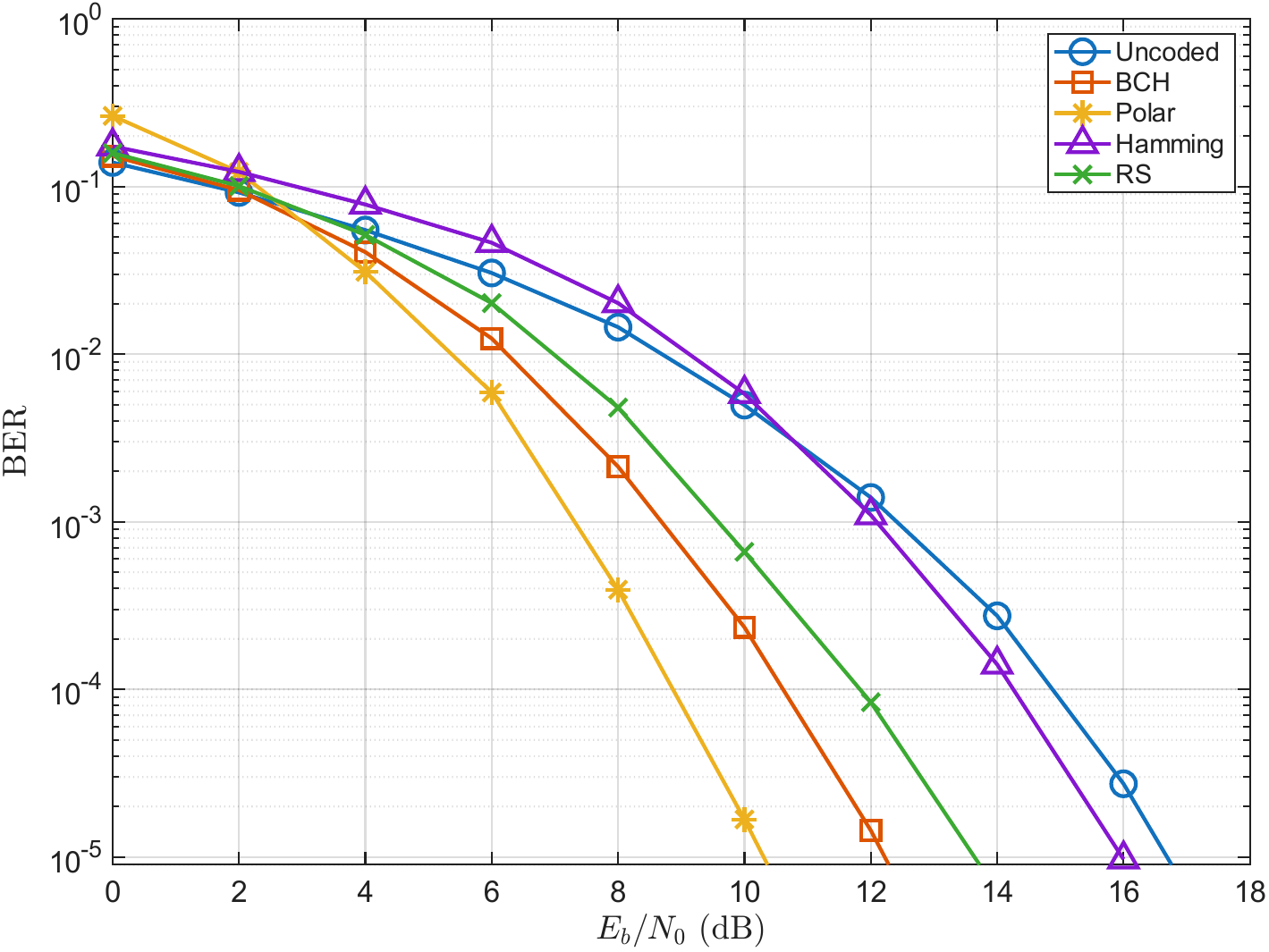}
    \caption{BER performance comparison of the CW-SVC/A$^2$FDM systems
      with BCH, polar, Hamming, RS, or no CLE.}
    \label{fig: channel coding}
\end{figure}

\fref{fig: channel coding} compares the BER performances of the
CW-SVC/A$^2$FDM systems aided by BCH, polar, Hamming or RS CLE or no
CLE. In addition to the CWC, the detailed parameter settings for the
CLEs are given as follows:
\begin{itemize}
    \item Uncoded: BPSK modulation, no CLE.
    \item BCH code: 4-QAM, BCH $(63,36)$, which can correct up to 5
      bit errors.
    \item polar code: 4-QAM, 32 information bits are applied with an 11-bit cyclic redundancy check (CRC) code. The length of the mother code is 64 bits. The mother code is decoded at the receiver using a CRC-aided successive cancellation list (CA-SCL) decoding scheme with a 16 bits list.
    \item Hamming code: 4-QAM, Hamming $(63,57)$, which can only
      correct 1 bit error.
    \item RS code: 4-QAM, RS $(14,8)$ over $\mathbb{GF}(2^4)$, which
      can correct up to 3 symbol errors.
\end{itemize}
Note that since different coding schemes require specific lengths of
input bits, we uniformly adopt zero padding at the end of data $\bm
b_\mathrm{cle}$ to adjust the input data length for a fair
comparison. The encoded bits are also padded with zeros to reach
$B_\mathrm{s}=64$ bits for QAM modulation. At the receiver, the padded
zeros are removed before and after decoding to recover the detected
data $\hat{\bm b}_\mathrm{cle}$. In the BER calculation, only the
original $B_\mathrm{cle}=32$ bits are taken into account, while the
zero padding bits are excluded in counting the errors.

As expected, as shown in \fref{fig: channel coding}, the uncoded case
exhibits the worst performance when a BER below $10^{-6}$ is demanded,
despite it employs the BPSK modulation of the lowest order. As only
one bit error can be corrected, the performance of the Hamming coded
scheme is the second worst one, offering only about 0.5 dB performance
gain at the BER of $10^{-5}$ over the uncoded system. By contrast, the
RS coded, BCH coded and polar coded scheme all achieve significant
performance gain in comparison with the Hamming coded
scheme. Specifically, among these three coded schemes, due to its
symbol-level error correction, RS coded scheme shows the worse
bit-level performance, but still provides a SNR gain of about 3 dB
over the uncoded case at the BER of $10^{-5}$. The BCH coded scheme,
which offers 5-bit error correction, results in a performance gain of
about 4.5 dB compared to the uncoded system. Among
  the coding schemes and parameter configurations evaluated in our
  simulations, the best performance is achieved by the polar coded
  scheme, which, aided by the CRC coding and the CA-SCL decoding,
  enables a 6.5 dB performance gain over the uncoded system.

\begin{figure}
    \centering
    \includegraphics[width=0.85\linewidth]{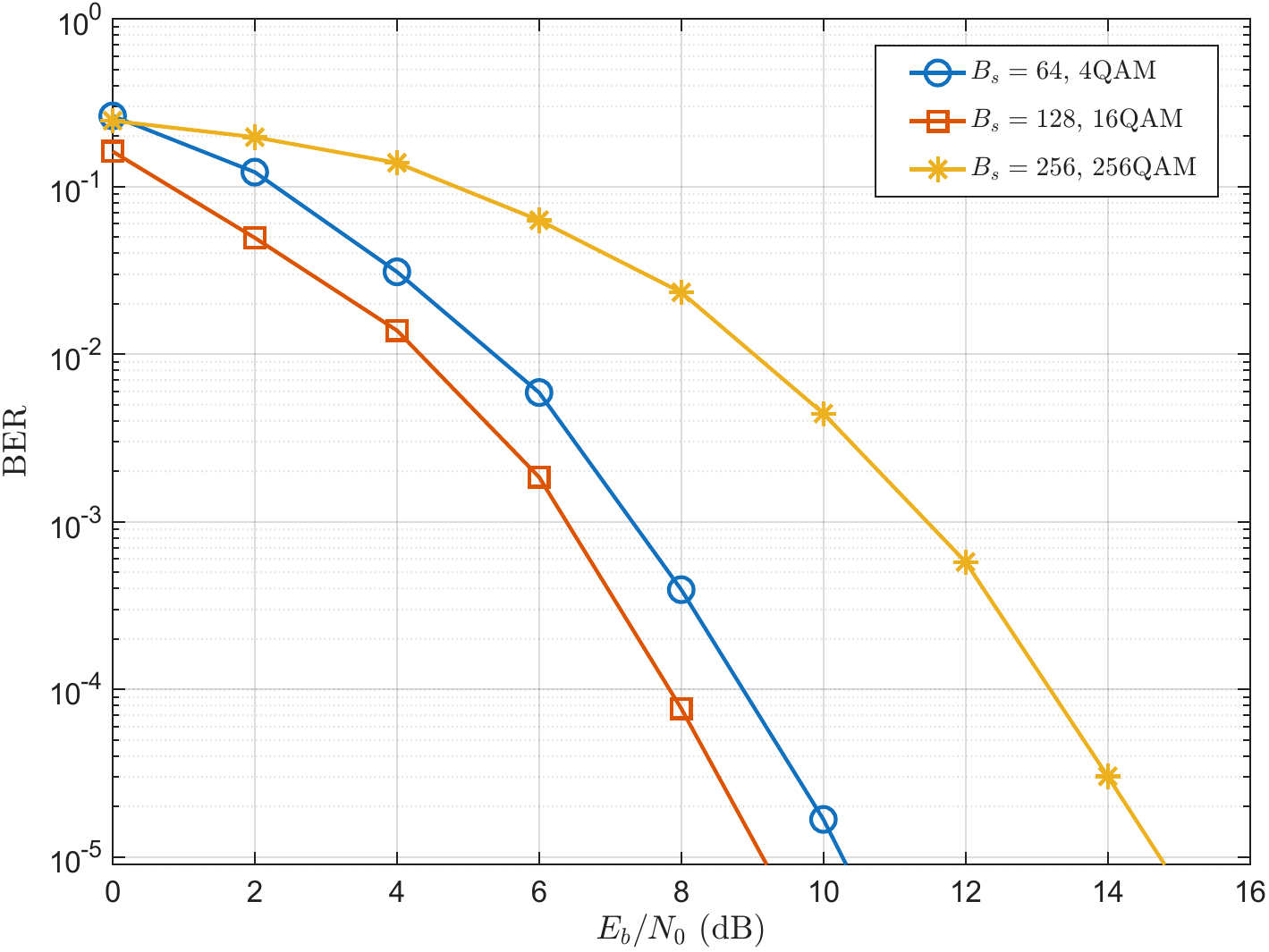}
    \caption{BER performance of the polar coded CW-SVC/A$^2$FDM
      systems with $B_\mathrm{cle}=32$ and $B_\mathrm{s} = 64,128$ and
      $256$.}
    \label{fig: Bs}
\end{figure}

Unlike BCH, Hamming and RS codes studied in \fref{fig: channel
  coding}, the error correction capability of polar codes cannot
generally be characterized by the fixed number of correctable bit or
symbol errors. \fref{fig: Bs} compares the BER performance of the
polar coded CW-SVC/A$^2$FDM systems with $B_\mathrm{cle}=32$ using
$B_\mathrm{s} = 64,128$ and $256$. For the polar coding
and decoding, $11$-bit CRC code and $16$-bit CA-SCL are
employed. Thus, the length of the input data of the polar encoder is always
$43$ bits. With this fixed number of input bits, a larger
$B_\mathrm{s}$ yields a lower coding rate, but requires a higher QAM
modulation to transmit the total code bits on the same number of
supports. Hence, as seen in \fref{fig: Bs}, 4QAM, 16QAM and 256QAM are
employed for the cases of $B_s=64$, $128$ and $256$, respectively.  It
is well-known that for the polar codes of the same length, a smaller
coding rate offers a higher capability to mitigate the effects of
noise and fading. On the other hand, given the same transmit power
determined by the number of information bits, a QAM of higher order
yields poorer reliability of detection than a QAM of lower
order. Hence, as shown in \fref{fig: Bs}, the performance of the
system with $B_\mathrm{s}=128$ (16QAM) is better than that of the
system with $B_\mathrm{s}=64$ (4QAM), with a gain of about $1$ dB.
However, the performance of the system with $B_\mathrm{s}=256$
(256QAM) is significantly worse than that of the system with either
$B_\mathrm{s}=64$ (4QAM) at a loss of 4.5~dB or $128$ (16QAM) at a
loss of $5.5$~dB. Hence, from the results we can conclude that for a
given information rate and the given number of supports, there is an
optimum selection for $B_s$ and the order of QAM, which enables a
polar coded CW-SVC/A$^2$FDM system to achieve the lowest BER within
some given SNR range.

\begin{figure}
    \centering
    \includegraphics[width=0.85\linewidth]{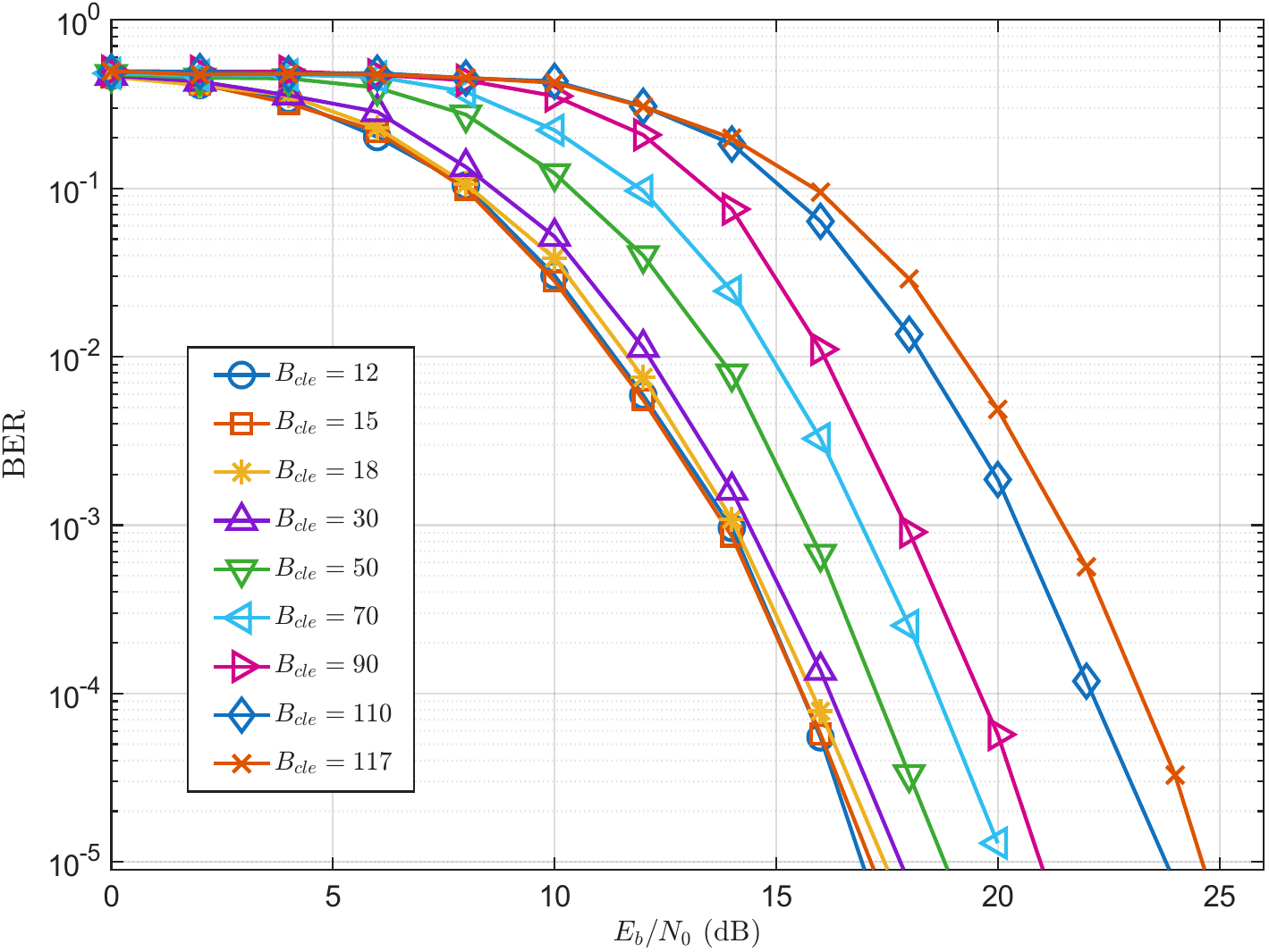}
    \caption{BER performance of the polar coded CW-SVC/A$^2$FDM
      systems with $B_\mathrm{s}=128$ for communications at different
      information rates of $B_\mathrm{cle} =12,15,18,30,50,70,90,110$
      and $117$.}
    \label{fig: Bcle}
\end{figure}

\fref{fig: Bcle} demonstrates the BER performance of the proposed system with a fixed length $B_\mathrm{s}=128$ of the polar-coded bits, but different values of $B_\mathrm{cle}=12,15,18,30,50,70,90,110$ and $117$, respectively. According to \cite{3gpp38.212}, a 6-bit CRC is employed for $B_\mathrm{cle}=12,15$ and $18$, while an 11-bit CRC is employed for the other cases. To ensure a fair comparison in terms of energy consumption, the transmitted signals are normalized such that the average energy per effective information bit is identical for all the considered values of $B_\mathrm{cle}$. Hence, for a fixed polar-coded length of $B_\mathrm{s}$, increasing $B_\mathrm{cle}$ yields a higher coding rate and a lower redundancy, thereby reducing the error-correction capability. Therefore, as shown in \fref{fig: Bcle}, the BER performance becomes worse as $B_\mathrm{cle}$ increases. The results verify that, given the energy per effective information bit, the decrease of coding redundancy has a significant impact on the BER performance. In the limit case of $B_\mathrm{cle}=117$, the $117$ information bits and 11 CRC bits occupy all the 128 inputs of the polar encoder, resulting in a polar code without frozen bits. Consequently, this configuration exhibits the poorest BER performance among all the considered cases.

\begin{figure}
    \centering
    \includegraphics[width=0.85\linewidth]{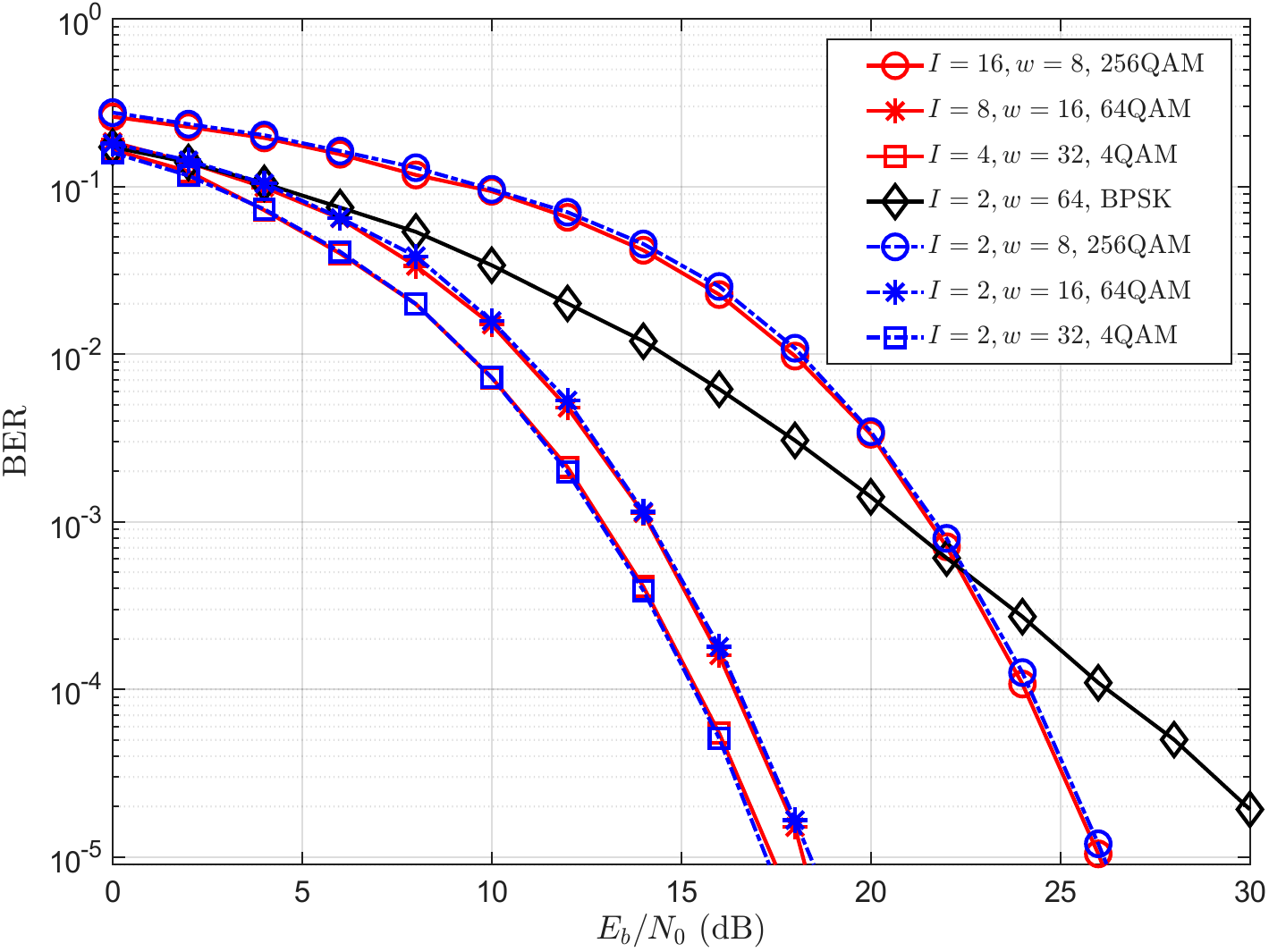}
    \caption{BER performance of the CW-SVC/A$^2$FDM systems employing
      the CW-codewords with supports of $w=8,16,32,64$ and 256 QAM, 64 QAM, 4
      QAM and BPSK modulation, respectively.}
    \label{fig: tradeoff}
\end{figure}

\fref{fig: tradeoff} compare the BER performances of the
CW-SVC/A$^2$FDM systems with different CWC codebooks, including
CWC$(128,128,64)$, CWC$(128,64,32)$, CWC$(128,32,16)$ and
CWC$(128,16,8)$. Accordingly, for a similar throughput, BPSK, 4 QAM,
64 QAM and 256 QAM are used for modulation. No CLE is considered to
avoid the effect from channel coding. There are two cases considered
in \fref{fig: tradeoff}. In the first case, the full codebooks are
utilized for SVC selection, which are marked as the red curves in
\fref{fig: tradeoff}, corresponding to $I=4$, $8$ and $16$. Here, the
numbers of candidate patterns $I$ are different, caused by the
different minimum Hamming distances $d_{min}$ of the CWC
considered. In the second case, the number of candidate patterns $I=2$
is chosen for all codebooks regardless of the CWC considered, which
are represented by the blue curves in \fref{fig: tradeoff}. From the
results of \fref{fig: tradeoff} it can be observed that the number of
candidate patterns has no effect on the performance of
systems. However, the performance improves as the weight $w$ increases
from $8$ to $16$ and then to $32$. This is because the minimum Hamming
distance increases by following the relation of $d_{min} =
2w$. Moreover, a larger weight means a big number of supports and
hence, more QAM symbols, which in turn means a lower modulation
order for transmitting the fixed number of information bits, hence
leading to performance improvement. However, when the weight is
increased to $w=64$, although the minimum Hamming distance is increased
to $d_{\min}=128$ and the modulation order is BPSK, the number of
supports per sparse vector becomes too large to be recovered from the
observed signal with a dimension of $N=64$. Consequently, its resulted
performance may be poor, especially, in the high SNR region where
interference dominates the achievable performance.

\begin{figure}
    \centering
    \includegraphics[width=0.85\linewidth]{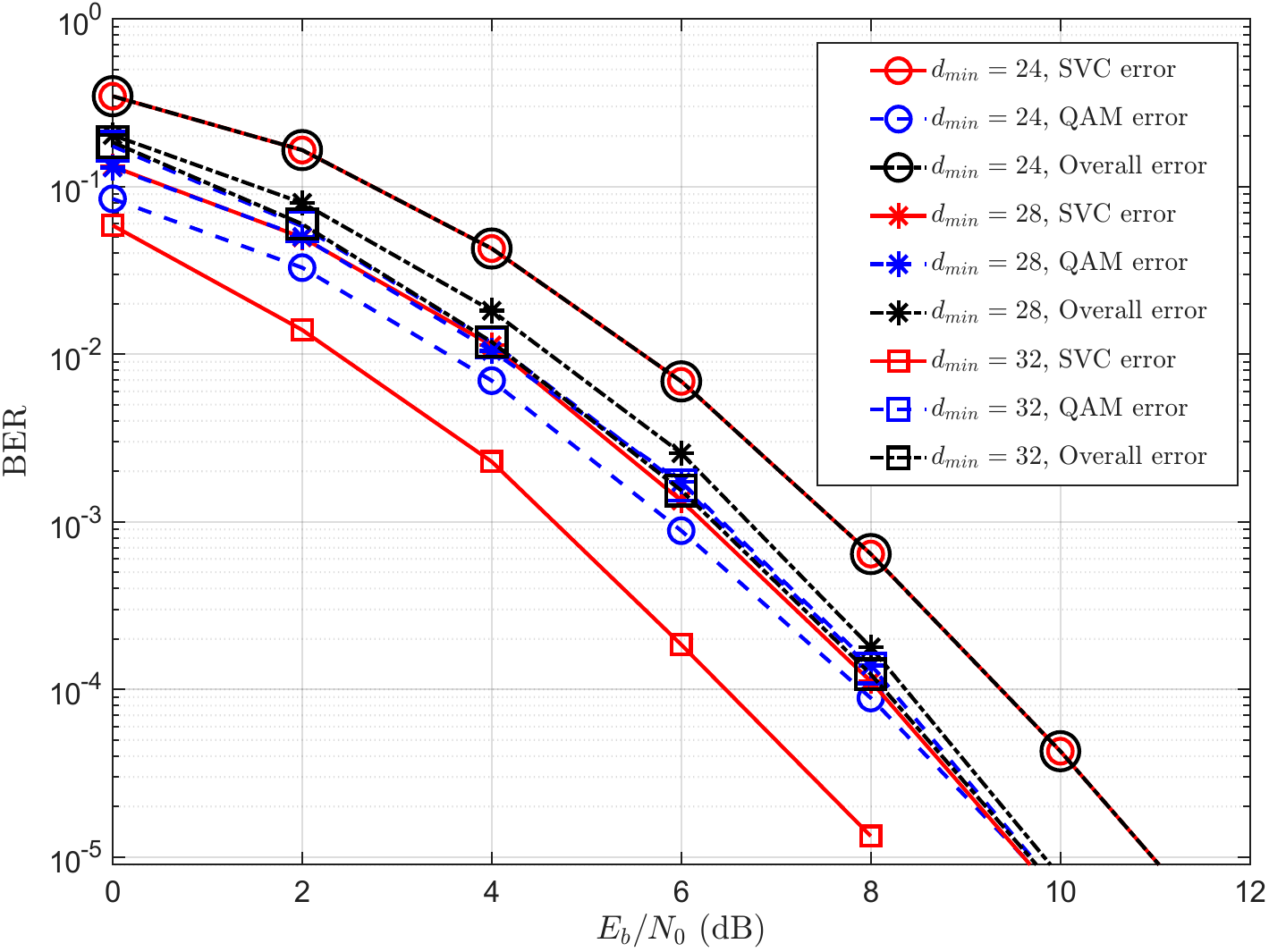}
    \caption{BER performance of the CW-SVC/A$^2$FDM systems where the
      CWCs with $w=16$ and $d_{min}=24$, $28$, and 32, respectively,
      are considered.}
    \label{fig: distance}
\end{figure}

\begin{table}
    \centering
\caption{Number of codewords in the codebooks considered in \fref{fig: distance}}
\label{tab:distance}
    \begin{tabular}{|c|c|c|c|}\hline
         CWC &  $t$ in FDTW&  $I_\mathrm{fdtw}$&  $I$\\\hline
         CWC$(128,32,16)$&  -&  -&  8\\\hline
         CWC$(128,28,16)$&  3&  72&  64\\\hline
         CWC$(128,24,16)$&  2&  2328& 2048\\ \hline
    \end{tabular}
\end{table}

In \fref{fig: distance}, the BER performances of the polar coded
CW-SVC/A$^2$FDM systems with three different CWCs are
demonstrated. The CWCs considered include CWC$(128,32,16)$,
CWC$(128,28,16)$ and CWC$(128,24,16)$. Among these CWCs,
CWC$(128,28,16)$ and CWC$(128,24,16)$ are generated by the FDTW
construction as detailed in Section \ref{section-ESVC-CWC design},
with $t=3$ and $t=2$, respectively, whereas CWC$(128,32,16)$ is
constructed directly with $d_{min}=2w$ as described in that
section. As shown in \tref{tab:distance}, the numbers of codewords of
these codebooks differ significantly. For the three codebooks,
$B_\mathrm{svc}$ equals $3$, $6$ and $11$, respectively, so that $I =
2^{B_\mathrm{svc}}$ support patterns are employed, namely $I = 8$,
$64$ and $2048$. As the results in \fref{fig: distance} show, the
overall error performance becomes better with the increase of
$d_{min}$. However, the situations for the systems with different
codebooks are different. Specifically, when $d_{min}=32$, the system
errors are dominated by the QAM/PSK demodulation and its followed CLE
decoding, the same as the previous results shown in \fref{fig:
  tradeoff}. When $d_{min}=28$, the number of codewords of a codebook
increases, resulting in the increase of the number of SVC detection
errors, which is close to the number of errors by QAM
demodulation. Accordingly, the overall BER is approximately equally
contributed by both the SVC errors and QAM/CLE errors. In the third
case when $d_{min}=24$, the decrease of minimum distance leads to
significant increase of the number of codewords, as shown in
\tref{tab:distance}, where the codebook of the CWC$(128,24,16)$
contains $I_\mathrm{fdtw}=2328$ codewords, of which
$I=2^{B_\mathrm{svc}}=2048$ are employed as the candidate support
patterns. In this case, the CW-SVC scheme converges to the
conventional ESVC scheme. Accordingly, the SVC errors overrun the
QAM/CLE errors, making the overall BER performance dominated by the
SVC errors.  Note again that although in this case, the overall BER
performance is still promising, about 1~dB worse than the other two
cases, the complexity of SVC detection is extremely high due to the
large number of codewords involved. Moreover, in our
  simulations, the detection is performed in serial, making the
  processing time growing with the number of candidate support
  patterns, which becomes significant when $I=2048$. If parallel
  implementation is employed, the critical-path latency can be
  significantly reduced at the cost of hardware resources that scale
  with $I$.

\begin{figure}
    \centering
    \includegraphics[width=0.85\linewidth]{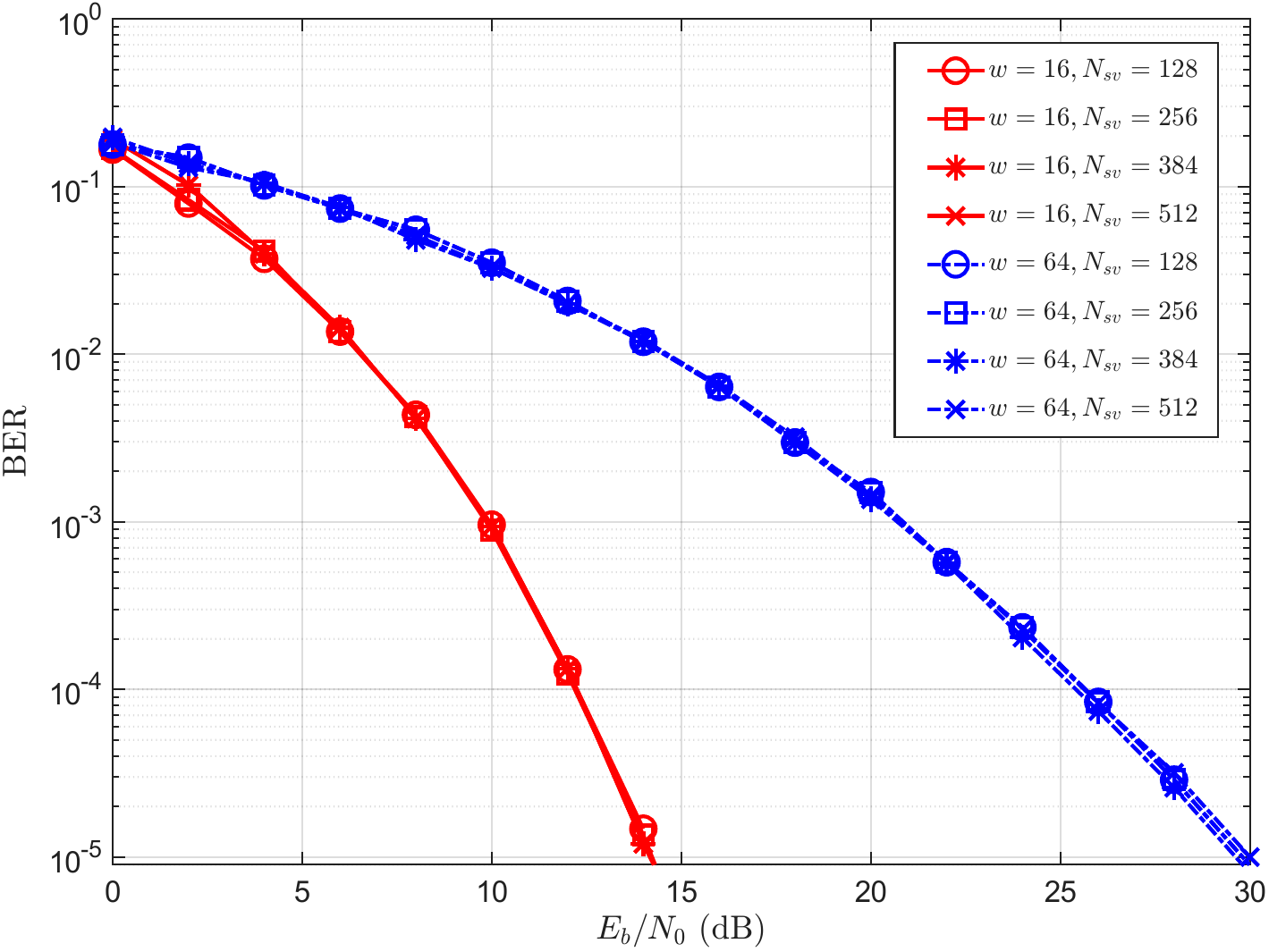}
    \caption{BER performances of the CW-SVC/A$^2$FDM systems with
      respectively $N_\mathrm{sv} =128,~256,~384,~512$ and
      $w=16,~64$.}
    \label{fig:CWC length}
\end{figure}

Finally, \fref{fig:CWC length} demonstrates the BER performance of the
uncoded CW-SVC/A$^2$FDM systems with the CWCs of different length of
$N_\mathrm{sv} = 128,~256,~384$ and $512$, while the
weight is $w=16$ or $64$. As shown in \fref{fig:CWC length}, for both $w=16$ and
$64$, the code length of $N_\mathrm{sv}$ yields no effect on the
system's BER performance, which is in contrast to the conventional ESVC
systems. The reason is that the number of candidates of support
patterns in CW-SVC/A$^2$FDM system is not directly related to the
length of the sparse vectors. As shown in \eqref{eq received signal
  jd}, the SVC detection only consider the supports in the sparse
vectors, i.e. the weight of CWC. As long as the number of supports,
i.e. weight, is fixed, the codeword length of $N_\mathrm{sv}$ does
not affect the system BER performance.

Compared with the conventional ESVC scheme, the
  CW-SVC without CLE achieves the BER level of $10^{-5}$ at an SNR
  about 14 dB lower, and the polar-coded CW-SVC further reduces the
  required SNR by about 6 dB. Under the same information throughput
  and transmission resources, the polar-coded scheme provides the
  strongest QAM protection, offering about a 6.5 dB gain over the
  uncoded case at the BER level of $10^{-5}$, outperforming the
  Hamming, RS and BCH coded schemes. For the polar-coded CW-SVC,
  $B_\mathrm{s}=128$ with 16QAM offers about a 1 dB gain over
  $B_\mathrm{s}=64$ with 4QAM, while $B_\mathrm{s}=256$ with 256QAM is
  clearly inferior. Within the reliably recoverable range, increasing
  the CWC weight and minimum Hamming distance improves the BER
  performance, whereas the sparse-vector length has no effect on the
  BER for a fixed weight.

\section{Conclusion} \label{section-ESVC-Conclusion}

In this paper, we proposed a CW-SVC/A$^{2}$FDM scheme that utilizes
CWC to protect the information conveyed by SVs, while employing CLE to
safeguard the QAM symbols transmitted on vector supports. This
architecture enables ultra-reliable low-latency short-packet
transmission in doubly selective fading environments. At the receiver,
a joint SV support and QAM detection algorithm was designed, which can
be efficiently implemented using parallel processing to reduce the
processing latency. Our theoretical analysis and simulation results
demonstrate that the proposed scheme yields a significant BER
improvement over conventional ESVC schemes while maintaining
low-latency parallel processing
feasibility. Accordingly, a low block error rate
  (BLER) in short-packet URLLC applications is possible, as BLER and
  BER is related by the union bound: $\mathrm{BLER}\leq
  B\cdot\mathrm{BER}$, where $B$ is the number of bits per packet. 
The results reveal that in CW-SVC/A$^{2}$FDM systems, CWC provides
ultra-reliable protection for SV-conveyed information, while
QAM-conveyed information is robustly protected by an appropriate
error-correction scheme. Evaluating a range of error-correction
candidates for QAM protection demonstrated that polar codes are the
most promising choice among the coding schemes and parameter
configurations evaluated in our simulations. Furthermore, the results
confirm that the CWC's weight and minimum Hamming distance must be
carefully selected to balance detection reliability, QAM modulation
order, and sparse recovery feasibility, thereby maximizing overall
system efficiency. Moving forward, our future research will extend
this framework to multi-user CW-SVC/A$^{2}$FDM systems operating in
diverse IoT environments, including mMTC and URLLC in device-centric
cell-free scenarios.

\begin{appendices}
  \section{constant weight codebooks}\label{Appendix-CWC}
  
There are six CWC codebooks considered in this paper, which are
$(128,128,64)$, $(128,64,32)$, $(128,32,16)$, $(128,16,8)$,
$(128,28,16)$ and $(128,24,16)$. CWC$(128,28,16)$ and
CWC$(128,24,16)$ are generated by the FDTW construction with $t=3$ and
$t=2$. All codewords in the remaining codebooks possess the same
property: their minimum Hamming distance satisfies
$d_{min}=2w$. Specifically, each codeword is assigned a distinct set
of indices in the sparse vector, such that all codewords in one
codebook have mutually different support pattern. The support indices
of the CWC$(128,32,16)$ are shown in \tref{tab:CWC(128,32,16)} as an
example.
\begin{table}[htbp]
    \centering
        \caption{CWC $(128,32,16)$}
    \label{tab:CWC(128,32,16)}
    \begin{tabular}{|c|c|c|c|c|c|c|c|}\hline
         $\bm \omega_0$&$\bm \omega_1$&$\bm \omega_2$&$\bm \omega_3$ & $\bm \omega_4$& $\bm \omega_5$& $\bm \omega_6$&$\bm \omega_7$\\\hline
         0&16&32&48 & 64& 80& 96&112\\\hline
         1&17&33&49 & 65& 81& 97&113\\\hline
         2&18&34&50 & 66& 82& 98&114\\\hline
         3&19&35&51 & 67& 83& 99&115\\\hline
         4&20&36&52 & 68& 84& 100&116\\\hline
         5&21&37&53 & 69& 85& 101&117\\\hline
         6&22&38&54 & 70& 86& 102&118\\\hline
         7&23&39&55 & 71& 87& 103&119\\\hline
         8&24&40&56 & 72& 88& 104&120\\ \hline
         9&25&41&57 & 73& 89& 105&121\\\hline
         10&26&42&58 & 74& 90& 106&122\\\hline
         11&27&43&59 & 75& 91& 107&123\\\hline
         12&28&44&60 & 76& 92& 108&124\\\hline
         13&29&45&61 & 77& 93& 109&125\\\hline
         14&30&46&62 & 78& 94& 110&126\\\hline
         15&31&47&63 & 79& 95& 111&127\\\hline
    \end{tabular}
\end{table}

\end{appendices}

\bibliography{reference}

\end{document}